\documentclass[]{mn2e}
\usepackage{epsfig}
\usepackage{graphicx}
\usepackage{mathtools}
\usepackage{pgf}

\title
{From evolved stars to the evolution of IC\,1613}
\author[S.A.\ Hashemi et al.]{
Seyed Azim Hashemi$^{1,2}$, Atefeh Javadi$^2$, Jacco Th. van Loon$^3$\\
$^1$Department of Physics, Sharif University of Technology, Tehran,
    11155-9161, Iran\\
$^2$School of Astronomy, Institute for Research in Fundamental Sciences (IPM),
    Tehran, 19395-5531, Iran\\
$^3$Lennard-Jones Laboratories, Keele University, ST5 5BG, UK}
\pagerange{\pageref{firstpage}--\pageref{lastpage}}
\pubyear{2018}
\begin{document}
\maketitle
\label{firstpage}
\begin{abstract}
IC\,1613 is a Local Group dwarf irregular galaxy at a distance of 750 kpc. In
this work, we present an analysis of the star formation history (SFH) of a
field of $\sim200$ square arcmin in the central part of the galaxy. To this
aim, we use a novel method based on the resolved population of more
highly evolved stars. We identify 53 such stars, 8 of which are supergiants
and the remainder are long period variables (LPV), large amplitude
variables (LAV) or extreme Asymptotic Giant Branch (x-AGB) stars. Using
stellar evolution models, we find the age and birth mass of these stars and
thus reconstruct the SFH. The average rate of star formation during
the last Gyr is $\sim3\times10^{-4}$ M$_\odot$ yr$^{-1}$ kpc$^{-2}$. The
absence of a dominant epoch of star formation over the past 5 Gyr, suggests
that IC\,1613 has evolved in isolation  for that long, spared
harrassment by other Local Group galaxies (in particular M\,31 and
the Milky Way). We confirm the radial age gradient, with star
formation currently concentrated in the central regions of IC\,1613, and the
failure of recent star formation to have created the main H\,{\sc i}
supershell. Based on the current rate of star formation at
$(5.5\pm2)\times10^{-3}$ M$_\odot$ yr$^{-1}$, the interstellar gas mass of the
galaxy of $9\times10^7$ M$_\odot$ and the gas production rate from AGB stars
at $\sim6\times10^{-4}$ M$_\odot$ yr$^{-1}$, we conclude that the star
formation activity of IC\,1613 can continue for $\sim18$ Gyr in a
closed-box model, but is likely to cease much earlier than that unless gas
can be accreted from outside.
\end{abstract}
\begin{keywords}
stars: AGB and post-AGB --
stars: supergiants --
stars: variables: general  --
galaxies: dwarf --
galaxies: evolution --
galaxies: individual: IC\,1613
\end{keywords}
\section{Introduction}
\label{sec:sec1}

The most abundant type of galaxy in the Local Group are low-mass, dwarf
galaxies and according to popular structure formation theory (e.g.,
White \& Rees 1978; Blumenthal et al.\ 1984; White \& Frenk 1991; Mo, van den
Bosch \& White 2010) we expect that this holds true for the whole Universe
 both in space and time. Their vicinity allows us access to
their structure, star formation history (SFH) and chemical composition, all
of which are the result of galaxy formation and evolution (e.g.,
Hodge 1989; Chun et al.\ 2015). Unlike their massive siblings, dwarf
galaxies are readily disturbed via interactions with larger galaxies
and the intra-cluster medium, as well as internal processes that remove gas
from these loosely bound systems (Stinson et al.\ 2007). Proximity
and simplicity of dwarf galaxies make them superb testcases to probe the
effect of different mechanisms operating in the internal and external
evolution of galaxies. Determining the SFH has a key role in this. In this
work, we set out to determine the SFH of IC\,1613, an isolated dwarf
irregular galaxy, and to use it to answer questions about its
interaction history, stellar population age gradient and morphology.

Much of the history of galaxies is imprinted upon the most highly
evolved stellar populations, spanning look-back times from $10^7$ to $10^{10}$
yr. The high luminosity of $\sim2000$ L$_\odot$ for tip-RGB stars,
$\sim10^4$ L$_\odot$ for Asymptotic Giant Branch (AGB) stars, and a few $10^5$
L$_\odot$ for red supergiants (RSGs) make cool evolved stars one of
the most accessible probes of the underlying stellar populations in the IR
(e.g., Maraston 2005; Maraston et al.\ 2006). Their spectral energy
distributions (SEDs) peak around 1 $\mu$m, so they stand out in the near-IR,
where extinction and reddening by dust is relatively low. They have low
surface gravity causing them to pulsate radially on timescales of a few weeks
to a few years. The Long Period Variable (LPV) stars are typically AGB stars
in the final stages of evolution (Fraser et al.\ 2005). All of the thermal
pulsing AGB (TP-AGB) stars are LPVs (e.g., Fraser et al.\ 2008;
Soszy\'nski et al.\ 2009), with periods of $>100$ days (e.g., Iben \&
Renzini 1983; Padova evolutionary tracks -- Marigo et al.\ 2008).
Furthermore, there is a good correlation between increasing period and
increasing amplitude (Fraser et al.\ 2008) and mass loss (Goldman et
al.\ (2017), and Large Amplitude Variable (LAV) stars therefore also bear a
strong relation to the end points of stellar evolution. In the absence of
period determinations, selection on the basis of amplitude would also result
in samples of the most highly evolved AGB stars.

In the AGB phase, the rate of mass loss accelerates with time. This is
because the luminosity and radius are increasing while the mass is
decreasing, thus leading to reduced surface gravity and less strongly bound
surface layers. In the final phase, a ``super wind'' develops characterised
by the highest dust fraction (e.g., Schr\"oder et al.\ 1999; Lagadec
\& Zijlstra 2008). Hence, very dusty AGB stars (extreme AGB stars, or x-AGB)
are at the end of their evolution on the AGB (e.g., Groenewegen et
al.\ 1998; Carroll \& Ostlie 2007).

While AGB stars originating from solar-mass stars have ages of $\sim10$ Gyr,
the most luminous AGB stars formed only $\sim10^8$ yr ago. To probe more
recently formed stellar populations, we include RSGs in our
analysis, which can be as young as $10^7$ yr old. The coolest RSGs also
pulsate with long periods and (in energy terms) considerable
amplitude (van Loon et al.\ 2008). We have developed a novel method
to use the relative numbers of these evolved AGB stars and
RSGs and their lifetimes to reconstruct the SFH (Javadi, van Loon \&
Mirtorabi 2011; Javadi et al.\ 2017; Rezaei Kh et al.\ 2014; Golshan et al.\
2017). In the following, we will apply this technique to IC\,1613.

IC\,1613 is an isolated dwarf galaxy within the Local Group, discovered by
Wolf (1906). We adopt the mean distance of 750 kpc ($(m -M)_0=24.37\pm0.08$
mag) from Menzies, Whitelock \& Feast (2015). Its vicinity, inclination angle
($i=38^\circ$; Lake \& Skillman 1989) and low foreground reddening
($E(B-V)=0.025$ mag; Schlegel, Finkbeiner \& Davis 1998) makes it a target of
choice for the study of its stellar populations. Its stellar mass is
estimated to be $\sim2\times10^8$ M$_\odot$ (Dekel \& Woo 2003; see
Orban et al.\ 2008), which is similar to its dynamical mass of
$(1.1\pm0.2)\times10^8$ M$_\odot$ estimated by Kirby et al.\ (2014) (the
observed maximum rotation velocity is 25 km s$^{-1}$; Lake \& Skillman
1989) suggesting no significant dark matter within the optical
half-light radius ($\sim1.4$ kpc).

The history of IC\,1613 is principally enshrined in its star formation
history. Cole et al.\ (1999) estimated a roughly constant star formation rate
(SFR) across the central 0.22 kpc$^2$ during the past 250--350 Myr, but 50\%
higher 400--900 Myr ago. The SFR in the central part over the past 300 Myr was
estimated by Bernard et al.\ (2007) to be  $(1.6\pm0.8)\times10^{-3}$ M$_\odot$
yr$^{-1}$ kpc$^{-2}$. Skillman et al.\ (2014), on the other hand, found
that the SFR in a small field near the half-light radius of IC\,1613 has been
constant over the entire lifetime of IC\,1613.

Chemical evolution places additional constraints on the history of a
galaxy, with the overall metallicity increasing in time as subsequent
generations of stars synthesize metals and return them to the interstellar
medium (ISM). Cole et al.\ (1999) found that the metallicity of IC\,1613 is
comparable to that of the Small Magellanic Cloud -- remarkable given
the latter is ten times more massive. Dolphin et al.\ (2001) derived a mean
value of [Fe/H] $=-1.15\pm0.2$ dex ($Z\sim0.001$); while Tikhonov \&
Galazutdinova (2002) derived a lower value of [Fe/H] $=-1.75\pm0.2$ dex
($Z\sim0.0003$) for the old population, the youngest population is expected to
be more metal rich. Indeed, Skillman, C\^ot\'e \& Miller (2003) found that
[Fe/H] has increased from $-1.3$ dex ($Z\sim0.0008$) at early times to $-0.7$
dex ($Z\sim0.003$) at present, which is confirmed by Tautvai\v{s}ien\.e et
al.\ (2007)'s estimate of [Fe/H] $=-0.67$ dex ($Z\sim0.003$) for the young
population.

The structure of this paper is as follows: in Section \ref{sec:sec2} we
describe the data we use. In section \ref{sec:sec3} we introduce our method.
Then, in section \ref{sec:sec4} we derive the SFH, followed by a discussion
and conclusions in sections \ref{sec:sec5} and \ref{sec:sec6}.

\section{Data}
\label{sec:sec2}

In this work we benefit from a number of published data sets at near- and
mid-IR wavelengths, which we shall describe in some detail below and which we
summarise in table~\ref{tab:tab1}.

\begin{table*}
\caption{Summary of photometric data.}
\label{tab:tab1}
\begin{tabular}{llcccccll}
\hline
Telescope     & Photometric      & Coverage    &
\multicolumn{4}{c}{\llap{---}-------------- number ---------------} &
Completeness Limit    & Reference               \\
              & bands       & (square arcmin) &
total        & AGB         & LPV & x-AGB &
(mag)                 &                         \\
\hline
IRSF          & J, H, K$_{\rm s}$ &         200 &
--           &         772 &  23 &   --  &
$K_{\rm s}=18.0$       & Menzies et al.\ (2015)  \\
{\it Spitzer} & [3.6], [4.5]     &         356 &
\llap{2}3538 & \llap{2}607 &  -- &   34  & 
$>75$\% to [3.6]=18.2 & Boyer et al.\ (2015a,b) \\
UKIRT         & J, H, K          & \llap{2}880 &
8624         &         843 &  -- &   --  &
$K=18.9$              & Sibbons et al.\ (2015)  \\
\hline
\end{tabular}
\end{table*}

\subsection{Near-infrared data}
\label{sec:sec2-1}

Menzies et al.\ (2015) published simultaneous JHK$_{\rm s}$ photometry from a
three-year monitoring survey with the 1.4-m InfraRed Survey Facility
of the central $\sim200$ square arcmin region of IC\,1613. They classified
all objects brighter than the tip of the first ascent red giant
branch (RGB; $K_{\rm s}=18$ mag) as supergiants or AGB stars (but not
foreground stars or background galaxies). They identified 758 variable stars
with a standard error $<0.1$ mag in the JHK$_{\rm s}$ bands, and 10
objects that had already been known in the literature to be supergiants.
Of these, 23 stars were identified as LAVs, for nine of which they
determined periods of $>100$ days.

Sibbons et al.\ (2015) used the Wide-Field CAMera on the 3.8-m UK
InfraRed Telescope to obtain JHK photometry of a wider, 0.8 square degree
area centered on $\alpha=1^{\rm h}4^{\rm m}54^{\rm s}$,
$\delta=2^{\circ}7^\prime57^{\prime\prime}$. Their catalogue presents de-reddened
photometry, listing 843 AGB stars within 4.5 kpc from the center of IC\,1613.
The tip of the RGB was determined at $K=18.25\pm0.15$ mag. From the colour
separation between the carbon (C) and M-type stars among the AGB population at
$(J-K)=1.15\pm0.05$ mag they determined a global C/M ratio of 0.52, and from
this [Fe/H] $=-1.26\pm0.08$ dex ($Z\sim0.0008$).

\subsection{Mid-infrared data}
\label{sec:sec2-2}

AGB stars are cool and produce dust during the thermal-pulsing phase. This
dust makes them appear redder, so longer wavelength data are needed to
identify the dustiest AGB stars that are in the final stages of evolution.
Boyer et al.\ (2015a) observed IC\,1613 as part of the DUST in Nearby
Galaxies Survey (DUSTiNGS), using the {\it Spitzer} Space Telescope at 3.6
and 4.5 $\mu$m to cover an area of 356 square arcmin on IC\,1613 in
two epochs, separated by 153 days.

In the Large Magellanic Cloud, Blum et al.\ (2006) classified stars with
$(J-[3.6])>3.1$ mag as ``extreme'' AGB or x-AGB stars. Boyer et al.\ (2015b)
devised a new criterion based solely on the {\it Spitzer} [3.6] and [4.5]
bands, with a 93--94\% success rate against using Blum's criterion in the
Magellanic Clouds. Hence, among the 50 new variable AGB candidates that Boyer
et al.\ detected, they identified 34 x-AGB candidates.

\subsection{Colour--magnitude diagrams}
\label{sec:sec2-3}

\begin{figure*}
\centerline{\hbox{
\epsfig{figure=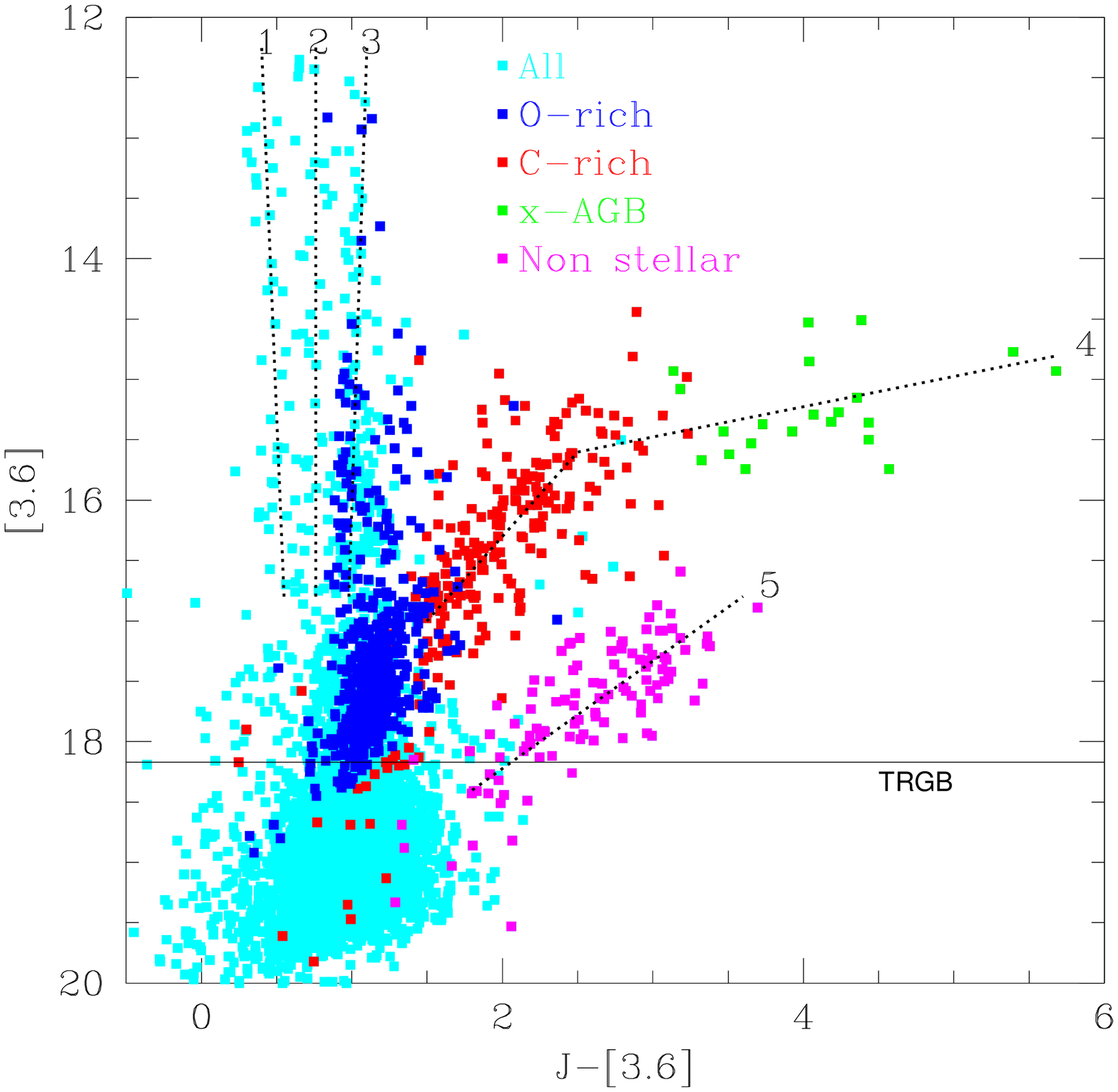,width=88mm}
\epsfig{figure=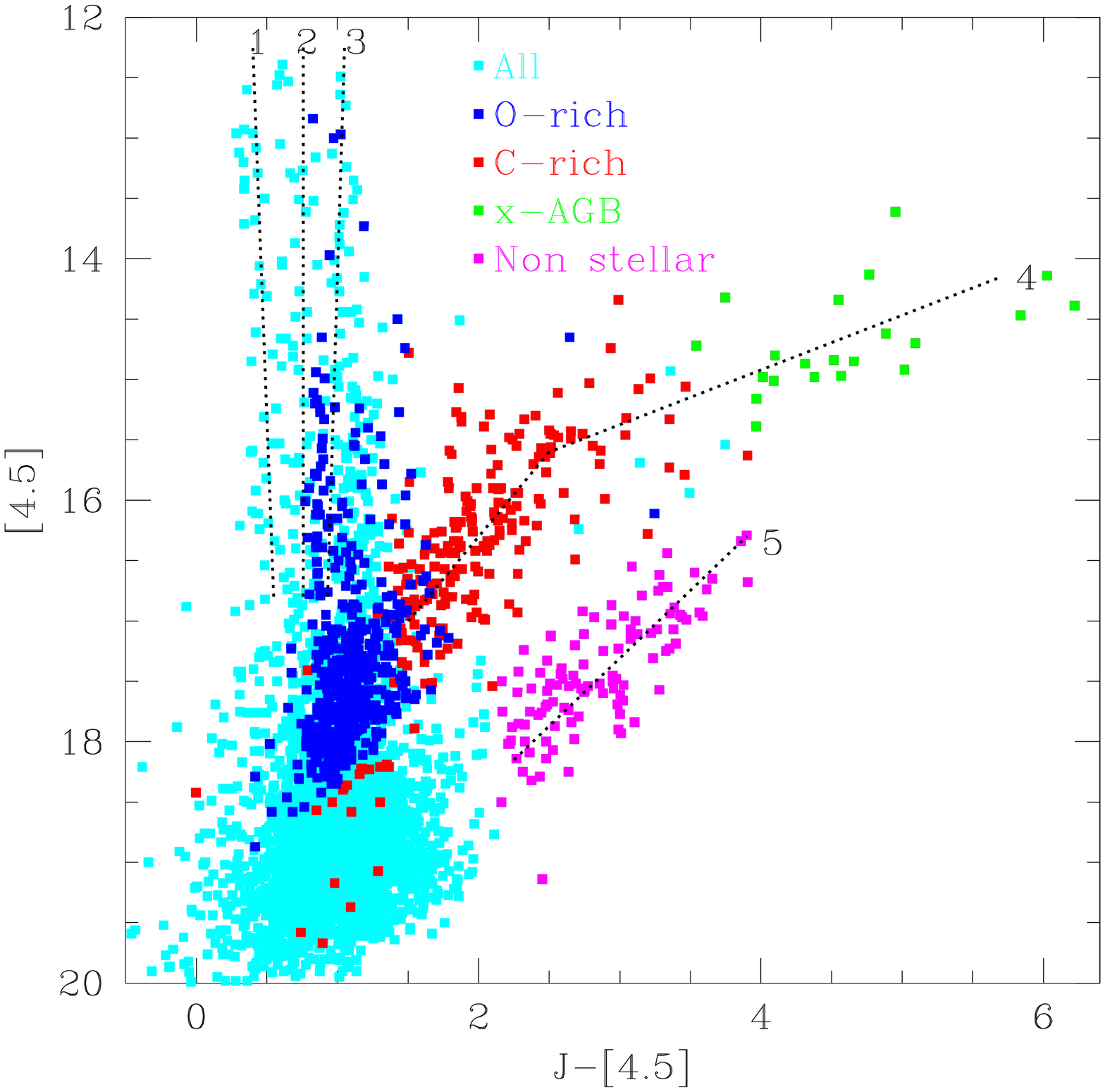,width=88mm}
}}
\caption[]{[3.6] vs.\ J--[3.6] ({\it left}) and [4.5] vs.\ J--[4.5] ({\it
right}) CMDs. All cross matched points are shown as cyan dots. Blue
dots are oxygen-rich (O-rich) stars, and red dots are carbon-rich (C-rich)
stars. The green dots represent x-AGB stars. The magenta dots are mainly
background galaxies (see text). The solid line in the left panel
marks the tip of the red giant branch (TRGB) at $[3.6]=18.17$ mag (Boyer et
al.\ 2009). Dashed numbered lines delineate sequences that we describe in the
text.}
\label{fig:fig1}
\end{figure*}

Here we examine CMDs, in order to arrive at the currently best
available sample of stars at their endpoints of evolution in the AGB
phase, that we can use to derive the SFH. We follow a similar identification
of features in the IR CMDs as in Blum et al.\ (2006), but using [4.5] instead
of [8].

We cross matched the stars from Sibbons et al.\ (2015) with stars in the
``good source catalogue'' from Boyer et al.\ (2015a), using a matching radius
of $1^{\prime\prime}$. We thus identified 5788 stars in common, of which 750
are AGB stars. About 30\% of Sibbons et al.'s sources were not
matched with any of Boyer et al.'s sources, which is largely a
result of the different spatial coverage.

The [3.6] vs.\ J--[3.6] and [4.5] vs.\ J--[4.5] CMDs are presented in
figure~\ref{fig:fig1}. Five sequences can be identified. The first of these,
sequence 1 ($J-[3.6]<0.5$ mag, reaching $[3.6]\sim12$ mag) corresponds to
young A--G-type supergiants. Separated by a few tenths of magnitude to the
red, sequence 2 consists mainly of foreground dwarfs and giants. Redder still
by a similar amount, sequence 3 corresponds to RGB stars, AGB stars and
late-type (mostly M) supergiants in IC\,1613 (Menzies et al.\ 2015;
Britavskiy et al.\ 2014; Herrero et al.\ 2010; Humphreys\ 1980).

Sibbons et al.\ (2015) divided the region above the tip of the RGB into
O-rich and C-rich where the latter have $(J-K)>1.15$ mag. In
figure~\ref{fig:fig1} the blue and red points correspond to Sibbons'
photometric division of O-rich and C-rich, respectively, where the division
occurs around $(J-[3.6])\sim1.4$ mag (similar for J--[4.5]). Of the 741 AGB
stars from Sibbons et al.\ that have $J$, $H$, $K$, [3.6] and [4.5], we
classed 477 ($\sim 64\%$) as O-rich and 264 ($\sim 36\%$) as C-rich.
This is very similar to Dell'Agli et al.\ (2016) where they compared
models to the combination of near-IR (Sibbons et al.\ 2015) and mid-IR (Boyer
et al.\ 2015a) data and found that the AGB sample of IC\,1613 is composed of
65\% O–rich and 35\% C-rich stars.

The C-stars (red) and x-AGB stars (green) form sequence 4. Up to
$(J-[3.6])\sim3.1$ mag ($(J-[4.5])\sim3.5$ mag) the sequence is dominated by
stars that were classified from near-IR photometry as C-rich; beyond that is
the realm of x-AGB stars. We stress that, while this sequence in other
galaxies is normally dominated by carbon stars (Zijlstra et al.\
2016; Wood et al.\ 2011) this does not preclude the presence of extremely
dusty O-rich AGB stars among these red sources (van Loon et al.\ 1997). We
classified all stars with $(J-[4.5])>3.5$ mag and $[4.5]<15.5$ mag as x-AGB;
these 21 objects are listed in table~\ref{tab:tab2}. Of these x-AGB stars,
all were found by Boyer et al.\ (2015b) to be variable (and x-AGB)
at mid-IR wavelengths except \#7773 and \#7091 (from the Sibbons et al.\
(2015) catalogue), though Menzies et al.\ (2015) did identify \#7773 as a
LAV.

\begin{table*}
\caption{Properties of the x-AGB stars identified in IC\,1613.}
\label{tab:tab2}
\begin{tabular}{ccccccccc}
\hline
RA  & Dec & Boyer & [3.6] & [4.5] & Sibbons & $J$ & $H$ & $K$ \\
(deg) & (deg) &    \# & (mag)   & (mag)   &      \# & (mag) & (mag) & (mag) \\
\hline
16.1832 & 2.0969 & \llap{1}42022 & $14.93\pm0.03$ & $14.32\pm0.03$ & \llap{1}1555 & $18.067\pm0.026$ & $16.786\pm0.015$ & $15.557\pm0.009$ \\
16.2993 & 2.1124 &         53171 & $15.35\pm0.02$ & $14.97\pm0.03$ & \llap{1}2772 & $19.536\pm0.099$ & $18.105\pm0.052$ & $16.936\pm0.030$ \\
16.2354 & 2.2274 &         99513 & $15.53\pm0.03$ & $14.87\pm0.03$ & \llap{1}9795 & $19.183\pm0.064$ & $17.871\pm0.038$ & $16.535\pm0.020$ \\
16.1821 & 2.0567 & \llap{1}42830 & $14.93\pm0.02$ & $14.39\pm0.03$ &         9057 & $20.610\pm0.234$ & $18.594\pm0.077$ & $16.917\pm0.029$ \\
16.1820 & 2.0887 & \llap{1}42987 & $15.29\pm0.03$ & $14.98\pm0.03$ & \llap{1}0922 & $19.359\pm0.080$ & $17.917\pm0.042$ & $16.764\pm0.026$ \\
16.0916 & 2.0912 & \llap{2}08444 & $15.43\pm0.03$ & $14.80\pm0.03$ & \llap{1}1103 & $18.900\pm0.054$ & $17.456\pm0.028$ & $16.312\pm0.017$ \\
16.2946 & 2.0223 &         56137 & $14.51\pm0.03$ & $14.13\pm0.03$ &         7773 & $18.896\pm0.058$ & $17.475\pm0.030$ & $16.132\pm0.015$ \\
16.2409 & 2.1547 &         95038 & $15.50\pm0.03$ & $14.92\pm0.03$ & \llap{1}5999 & $19.935\pm0.121$ & $18.438\pm0.062$ & $17.097\pm0.032$ \\
16.0911 & 2.2047 & \llap{2}08785 & $15.08\pm0.02$ & $14.72\pm0.03$ & \llap{1}8795 & $18.262\pm0.034$ & $17.030\pm0.021$ & $16.025\pm0.015$ \\
16.2147 & 2.0639 & \llap{1}15947 & $15.36\pm0.03$ & $14.70\pm0.03$ &         9409 & $19.794\pm0.118$ & $18.177\pm0.053$ & $16.666\pm0.024$ \\
16.1727 & 2.1928 & \llap{1}50523 & $15.74\pm0.03$ & $15.39\pm0.03$ & \llap{1}8237 & $19.354\pm0.086$ & $18.114\pm0.053$ & $17.033\pm0.035$ \\
16.3041 & 2.0853 &         50487 & $15.27\pm0.03$ & $14.62\pm0.03$ & \llap{1}0697 & $19.505\pm0.096$ & $18.224\pm0.058$ & $16.556\pm0.022$ \\
16.2899 & 2.2320 &         59151 & $14.53\pm0.03$ & $13.61\pm0.03$ & \llap{1}9976 & $18.563\pm0.037$ & $17.085\pm0.019$ & $15.611\pm0.009$ \\
16.2061 & 2.2790 & \llap{1}22963 & $15.62\pm0.02$ & $15.16\pm0.03$ & \llap{2}1513 & $19.129\pm0.071$ & $17.936\pm0.046$ & $16.785\pm0.028$ \\
16.1948 & 2.0021 & \llap{1}32310 & $14.77\pm0.02$ & $14.14\pm0.03$ &         7091 & $20.165\pm0.163$ & $18.349\pm0.062$ & $16.829\pm0.027$ \\
16.1777 & 2.1171 & \llap{1}46410 & $15.43\pm0.03$ & $14.84\pm0.03$ & \llap{1}3166 & $19.356\pm0.081$ & $17.872\pm0.041$ & $16.513\pm0.021$ \\
16.2195 & 2.1156 & \llap{1}12075 & $14.85\pm0.03$ & $14.34\pm0.03$ & \llap{1}3031 & $18.889\pm0.054$ & $17.537\pm0.030$ & $16.111\pm0.015$ \\
16.2480 & 2.2076 &         89380 & $15.37\pm0.02$ & $15.01\pm0.03$ & \llap{1}8925 & $19.100\pm0.059$ & $17.461\pm0.026$ & $16.545\pm0.020$ \\
16.2351 & 2.2527 &         99713 & $15.74\pm0.02$ & $14.47\pm0.03$ & \llap{2}0671 & $20.309\pm0.169$ & $19.625\pm0.178$ & $18.210\pm0.088$ \\
16.1770 & 2.1128 & \llap{1}46972 & $15.15\pm0.03$ & $14.85\pm0.03$ & \llap{1}2808 & $19.508\pm0.092$ & $18.031\pm0.047$ & $16.615\pm0.023$ \\
16.1880 & 2.1350 & \llap{1}38007 & $15.67\pm0.03$ & $14.98\pm0.03$ & \llap{1}4597 & $18.993\pm0.064$ & $17.724\pm0.038$ & $16.686\pm0.026$ \\
\hline
\end{tabular}
\end{table*}

Finally, sequence 5 (magenta in Fig.~\ref{fig:fig1}) are predominantly
background galaxies (Meixner et al.\ 2006).

The numbers, and their contributions, of various populations identified in the
CMDs are summarised in table~\ref{tab:tab3}.

\begin{table}
\caption{Summary of CMD selections of stellar populations
and colour plotted in Fig.~\ref{fig:fig1} .}
\label{tab:tab3}
\begin{tabular}{lcrcl}
\hline
Population              & Diagram &  $N$ &                  \% & Colour  \\
\hline
Sources with J \& [3.6] & [3.6], J--[3.6] & \llap{5}261 & \llap{10}0          & Cyan    \\
AGB with J \& [3.6]     & [3.6], J--[3.6] &  741 &  \llap{1}4          &                  \\
C-rich AGB              & [3.6], J--[3.6] &  264 &          5          & Red     \\
O-rich AGB              & [3.6], J--[3.6] &  477 &          8          & Blue    \\
x-AGB                   & [4.5], J--[4.5] &   21 &          0\rlap{.4} & Green   \\
Background galaxies     & [4.5], J--[4.5] &  115 &          2          & Magen\rlap{ta} \\
\hline
\end{tabular}
\end{table}

\subsection{Sample selection to determine the SFH}
\label{sec:sec2-4}

Previous application of our method (Javadi et al.\ 2011, 2017; Rezaei
Kh et al.\ 2014; Golshan et al.\ 2017) -- described in Section~\ref{sec:sec3}
-- was based on confirmed LPVs. LPVs are red giant or supergiant pulsating
stars with periods ranging from months to a few years (e.g., Soszy\'nski et
al.\ 2009). In the case at hand, though, the limited cadency of the DUSTiNGS
survey and the limited depth of the Menzies et al.\ (2015) survey will have
led to LPVs being missed. Therefore, we combine confirmed LPVs with those AGB
and RSG candidate stars that are also expected to be LPVs. These candidates
are:
\begin{enumerate}
\item x-AGB stars (a combination of our selection and Boyer et al.\
2015b) and LAVs without determined period that are expected to be LPVs near
the end of the AGB phase (e.g., Schr\"oder et al.\ 1999; Fraser et al.\ 2008;
Soszy\'nski et al.\ 2009).
\item RSGs that do not have a determined period but must be good
candidates for being LPVs. We note that any meaningful period determination of
RSGs may require decades of observations (e.g., Kiss et al.\ 2006; Pierce et
al.\ 2000).
\end{enumerate}

\begin{figure*}
\centerline{\hbox{
\epsfig{figure=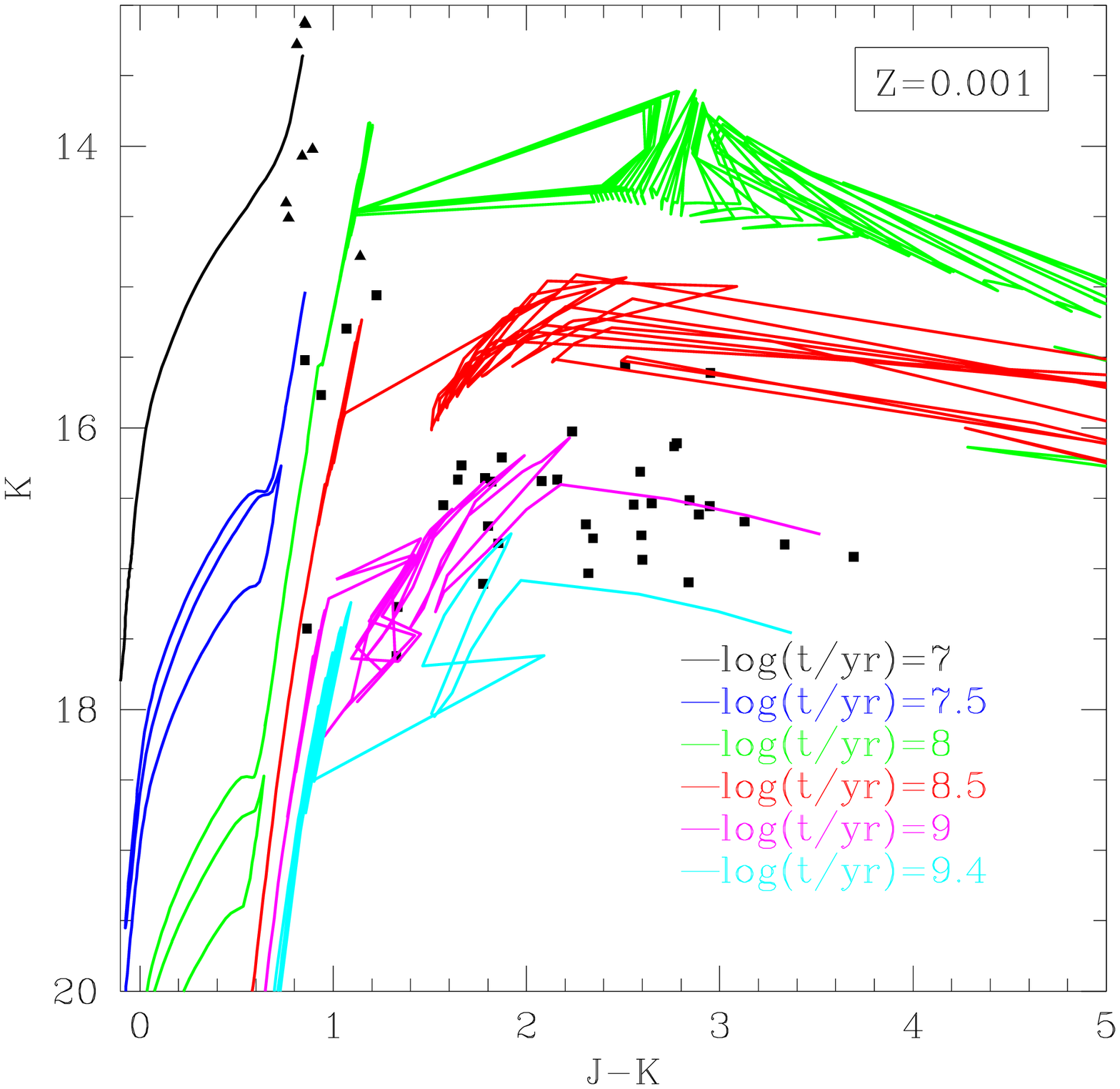,width=88mm}
\epsfig{figure=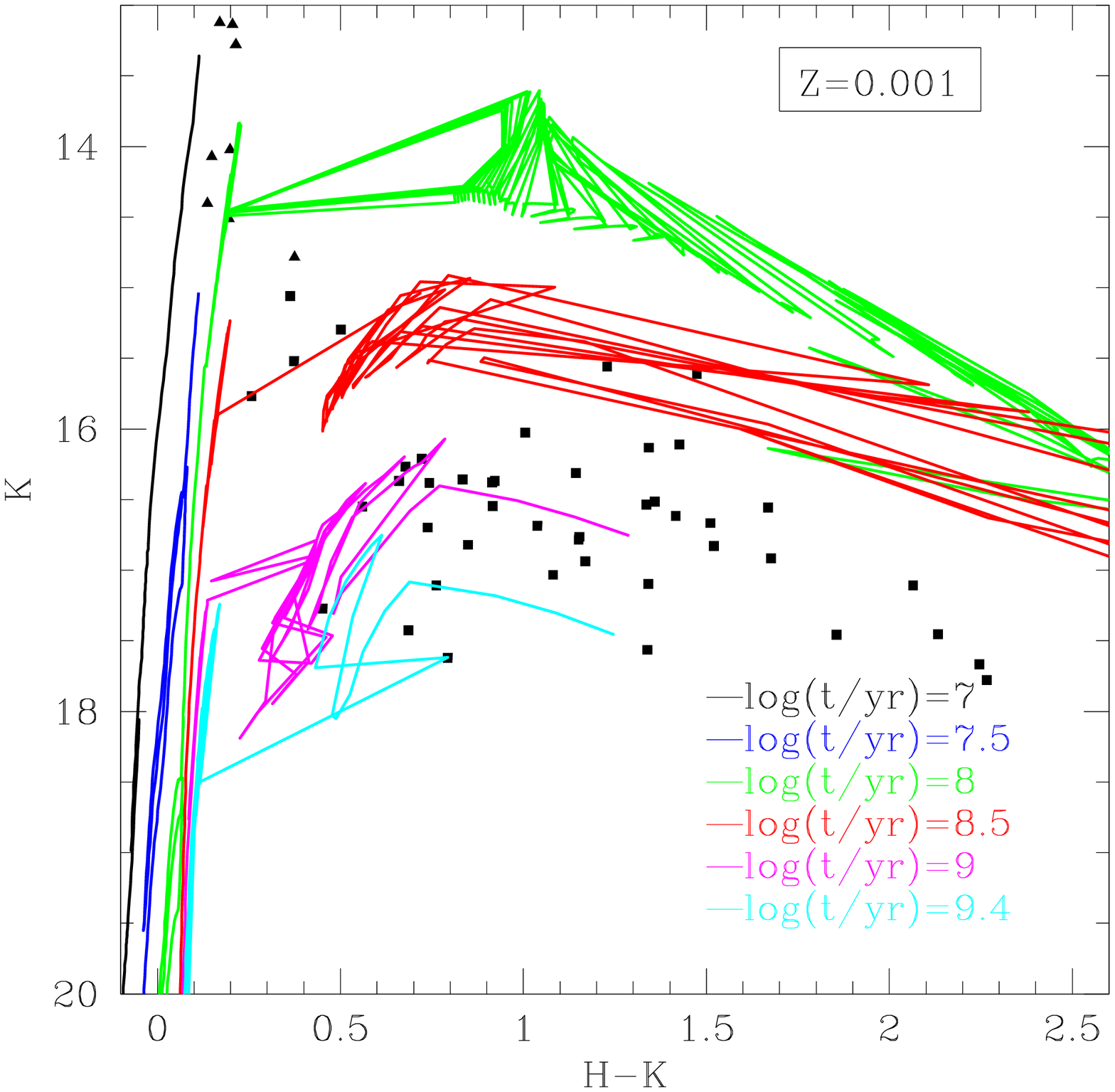,width=88mm}
}}
\caption[]{K vs.\ J--K ({\it left}) and H--K ({\it right}) with our selection
of evolved stars (squares: AGB LPVs; triangles: RSGs), overlain with
isochrones from Padova models (Marigo et al.\ 2017) labeled with logarithmic
ages.}
\label{fig:fig2}
\end{figure*}

Based on their location in the CMDs and with respect to isochrones
(Fig.~\ref{fig:fig2}; Padova models from Marigo et al.\ 2017), our sample
occupies the same region as the LPVs in our previous works.

The sample is summarised in table~\ref{tab:tab4}; of the x-AGB stars, eight
were identified as LPVs or LAVs, hence the number of unique sources is 53 (two
of the RSGs are also LPVs but they have not been counted in that category in
the table).

\begin{table}
\caption{Description of our sample to derive the SFH in IC\,1613 (Total excludes duplicates).}
\label{tab:tab4}
\begin{tabular}{lrl}
\hline
Population & $N$ & Reference                                             \\
\hline
LPV        &  14 & Menzies et al.\ (2015)                                \\
LAV        &   9 & Menzies et al.\ (2015)                                \\
x-AGB      &  30 & Boyer et al.\ (2015b) (cf.\ section~\ref{sec:sec2-3}) \\
RSG        &   8 & Menzies et al.\ (2015)                                \\
Total      &  53 &                                                       \\
\hline
\end{tabular}
\end{table}

\begin{figure*}
\centerline{\hbox{\epsfig{figure=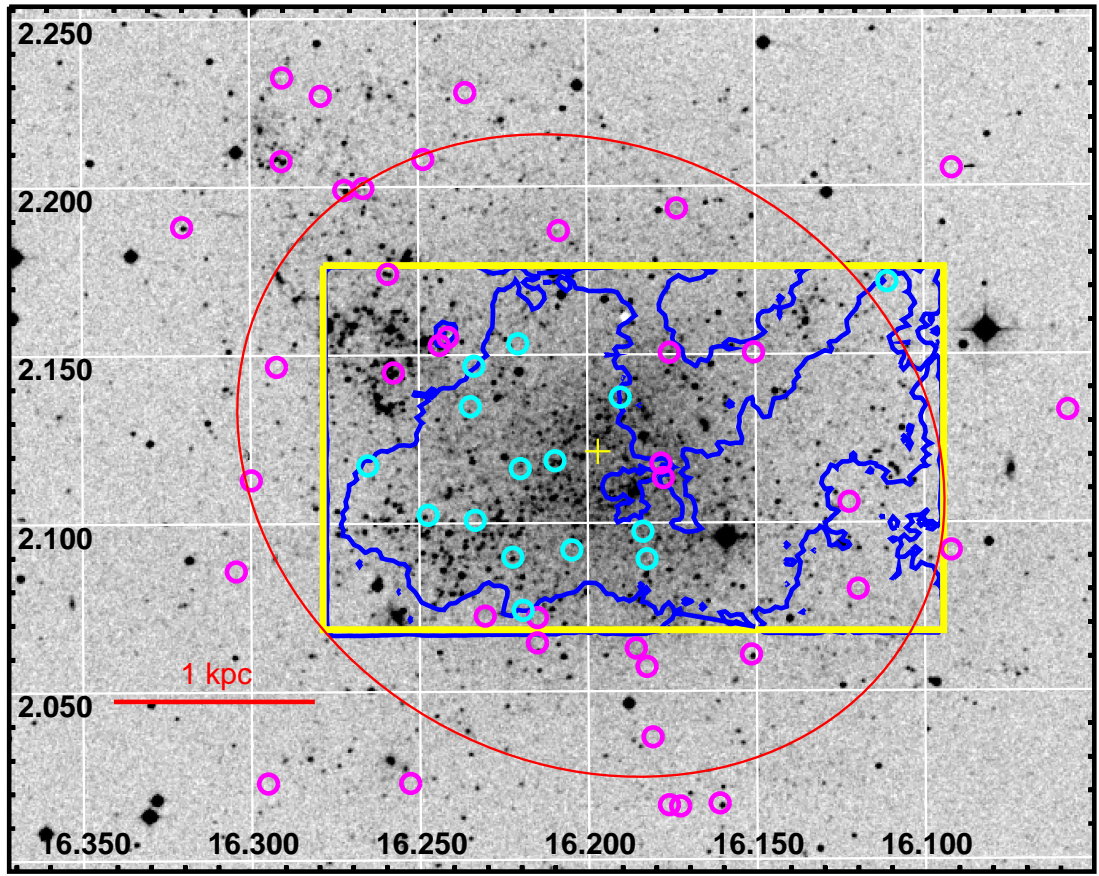,width=170mm}}}
\caption[]{Spatial distribution of our stellar sample (magenta and
cyan circles) and neutral hydrogen shells (H\,{\sc i}; the largest blue
contours in the yellow box). The stars of the sample which are inside the
approximate boundary of the main H\,{\sc i} supershell (Silich et al.\ 2006)
are shown with cyan circles. The yellow cross indicates the optical centre of
IC\,1613, with the red ellipse outlining the half-light radius
($r_{\rm h}=6.5^\prime$, position angle $PA=58^\circ$). The (archival) optical
image (at 468 nm wavelength) in the background was made with the UK Schmidt
Telescope.}
\label{fig:fig3}
\end{figure*}

We limit the field of study to the smaller of the two, Boyer et al.\ (2015a)
and Menzies et al.\ (2015), measuring 11.8 kpc$^2$ (after deprojection). Our
sample is shown in figure~\ref{fig:fig3}, in relation to the general
stellar distribution and the neutral hydrogen gas. As the half-light
radius suggests, as many sources could be expected to be found (just) outside
that radius as within that radius. While there is no direct confirmation of
membership for any of these sources, their concentration on the (small portion
of) sky strongly suggests that most -- if not all -- are associated with
IC\,1613.

\section{Method: deriving the SFH}
\label{sec:sec3}

The method we employ here was developed in
Javadi et al.\ (2011, 2017). We limit ourselves to a brief description along
with the model parametrisations determined for our case.

The SFH is the rate at which gas is converted into stars, $\xi$ (in M$_\odot$
yr$^{-1}$), as a function of time. The amount of stellar mass, $dM$, produced
in a time interval $dt$ is:
\begin{equation}
dM(t) = \xi(t)\ dt.
\label{eq:eq1}
\end{equation}
We can find the number of formed stars from the total produced stellar mass
by:
\begin{equation}
dN(t) = \frac{\int_{\rm min}^{\rm max}f_{\rm IMF}(m)\ dm}
{\int_{\rm min}^{\rm max}f_{\rm IMF}(m)m\ dm}\ dM(t),
\label{eq:eq2}
\end{equation} 
where $f_{\rm IMF}$ is the initial mass function (IMF). We use the IMF defined
in Kroupa (2001):
\begin{equation}
f_{\rm IMF} = Am^{-\alpha},
\label{eq:eq3}
\end{equation}
where $A$ is the normalisation coefficient and $\alpha$ is a factor which
depends on the mass range:
\begin{equation}
\alpha = \left\{\begin{array}{cclcl}
+0.3\pm0.7 & {\rm for} & {\rm min} & < \frac{m}{{\rm M}_\odot} < & 0.08 \\
+1.3\pm0.5 & {\rm for} & 0.08      & < \frac{m}{{\rm M}_\odot} < & 0.5 \\
+2.3\pm0.3 & {\rm for} & 0.5       & < \frac{m}{{\rm M}_\odot} < & {\rm max} \\
\end{array}\right.
\label{eq:eq4}
\end{equation}
The minimum and maximum mass in Kroupa's IMF were assumed to be 0.02 and 200
M$_\odot$, respectively.

Given that we use LPVs as proxies for the underlying stellar populations, we
need to relate the number of variable stars to the total number of stars, $N$,
which were formed in $dt$. We note that in this section `LPVs' stand
for all identified LPVs and other candidates that we expect to be in this
phase (cf.\ section \ref{sec:sec2-4}). If stars with mass between $m(t)$ and
$m(t+dt)$ ($t$ is look-back time) are now LPVs, then the number of LPVs
created between times $t$ and $t+dt$ is:
\begin{equation}
dn(t) = \frac{\int_{m(t)}^{m(t+dt)}f_{\rm IMF}(m)\ dm}
{\int_{\rm min}^{\rm max}f_{\rm IMF}(m)\ dm}\ dN(t).
\label{eq:eq5}
\end{equation}
By substituting equations~\ref{eq:eq1} and \ref{eq:eq2} into
equation~\ref{eq:eq5} we have:
\begin{equation}
dn(t) = \frac{\int_{m(t)}^{m(t+dt)}f_{\rm IMF}(m)\ dm}
{\int_{\rm min}^{\rm max}f_{\rm IMF}(m)m\ dm}\ \xi(t)\ dt.
\label{eq:eq6}
\end{equation}

The number of LPVs, $dn^\prime$, we can observe in an age bin, $dt$, depends
on the size of the bin ($dt$) and on the duration of the LPV stage ($\delta
t$):
\begin{equation}
dn^\prime(t) = \frac{\delta t}{dt}\ dn(t).
\label{eq:eq7}
\end{equation}

Finally, by combining the above equations, we obtain a relation to give the
SFR based on LPV counts:
\begin{equation}
\xi(t) = \frac{\int_{\rm min}^{\rm max}f_{\rm IMF}(m)m\ dm}
{\int_{m(t)}^{m(t+dt)}f_{\rm IMF}(m)\ dm}\ \frac{dn^\prime(t)}{\delta t}.
\label{eq:eq8}
\end{equation}

In order to relate a LPV's brightness to its mass, and its mass to its age
(look-back time $t$) we appeal to stellar evolution models. The most suitable
models are those from the Padova group, as argued extensively in Javadi et
al.\ (2011, 2017); Rezaei Kh et al.\ (2014); Golshan et al.\ (2017). Here we
use the latest version (PARSEC v1.2S + COLIBRI PR16; Marigo et al.\ 2017),
which was improved upon the previous one (Marigo et al.\ 2008) resulting
in a better estimation of the birth mass and pulsation timescale. Some cases of improvements that matter for us here are: new TP-AGB evolutionary tracks and atmosphere
models for O-rich and C-rich stars; the complete thermal pulse cycles, with a
full description of the in-cycle changes in the stellar parameters; new
pulsation models to describe the fundamental and first overtone modes of
LPVs; new dust models that consider the growth of the grains during the AGB
evolution. That said, the evolutionary tracks of the youngest (most massive)
stars ($\log M/{\rm M}_\odot \ga 0.7$) were not updated and still hail from the 2008 publication.

The LPVs are assumed to have reached their maximum near-IR brightness. Hence
we apply a mass--K-band magnitude relation appropriate for the metallicity
and distance modulus of IC\,1613 ($Z=0.001$ and $\mu=24.37$ mag; left panel
of figure~\ref{fig:fig4}; all the figures and tables for other
metallicities we use in this paper are presented in the Appendix).
While the K-band magnitude varies during the radial pulsation cycle,
the photometry we are using are the mean magnitudes over several epochs and
thus representative of the luminosity associated with the nuclear burning
inside the star. The evolution of super-AGB stars toward the brighter
K-band magnitudes has been omitted from the models, thus we interpolate the
models over that range in mass (see Javadi et al.\ 2011 for details). The
coefficients of the linear fitting between K-band magnitude and mass are
listed in table~\ref{tab:tab5}.

\begin{table}
\caption[]{Fitted equations for the relation between birth mass and K-band
magnitude, $\log M/{\rm M}_\odot=aK+b$, for a distance modulus of $\mu=24.37$
mag.}
\begin{tabular}{ccr}
\hline\hline
\multicolumn{3}{c}{$Z=0.001$} \\
\hline
       $a$       &      $b$         &   validity range   \\
\hline
$-0.322\pm0.009$ &  $5.560\pm0.168$ & $K\leq13.20$       \\
$-1.726\pm0.051$ & $24.19\pm0.720$ & $13.20<K\leq13.61$  \\
$-0.176\pm0.005$ &  $3.092\pm0.009$ & $13.61<K\leq14.13$ \\
$-0.072\pm0.002$ &  $1.617\pm0.048$ & $14.13<K\leq14.74$ \\
$-0.339\pm0.016$ &  $5.552\pm0.168$ & $14.74<K\leq15.28$ \\
$-0.116\pm0.004$ &  $2.140\pm0.063$ & $15.28<K\leq16.07$ \\
$-0.175\pm0.006$ &  $3.090\pm0.009$ & $16.07<K\leq16.79$ \\
$-0.107\pm0.003$ &  $1.944\pm0.057$ & $16.79<K\leq17.40$ \\
$-0.225\pm0.006$ &  $4.008\pm0.120$ & $K>17.40$          \\
\hline
\end{tabular}
\label{tab:tab5}
\end{table}

The onset of dust formation reddens and dims the stars, and we therefore need
to apply a correction to bring them back to their photospheric peak
brightness level. Because the evolutiuonary timescale shortens
dramatically once dust formation sets in, the same brightness level found in
the model at the end point of stellar evolution corresponds to the nuclear
burning luminosity that can be directly related to the birth mass. Because
some of the stars in our sample do not have J-band detections, we use H--K
colours to determine this correction. As clearly discernible in
figure~\ref{fig:fig2}, the reddened parts of the tracks/isochrones
are bimodal in terms of their slope, based on whether the dust is O-rich or
C-rich (Menzies et al.\ 2015; Albert et al.\ 2000). Therefore, we
apply the following correction:
\begin{equation}
K _{\rm cor} = K+a_{\rm oxygen}[(H-K)-(H-K)_0],
\label{eq:eq9}
\end{equation}
for O-rich stars. For C-rich stars according to the isochrones in the right
panels of Fig. \ref{fig:fig2} we have:
\begin{equation}
K _{\rm cor} = K+a_{\rm carbon}[(H-K)-(H-K)_0],
\label{eq:eq10}
\end{equation}
where $a_{\rm oxygen}=-1.2$ and $a_{\rm carbon}=-0.85$ are the average slopes of
the isochrone after the peak in the K vs.\ H--K CMDs, and the average colours
at the peak brightness are $(H-K)_0=1.05$ mag for O-rich stars and
$(H-K)_0=0.8$ mag for C-rich stars. In some cases there is
spectroscopic evidence for the chemical class of the object (Menzies et al.\
2015; Albert et al.\ 2000); where this is not the case we have relied on
photometric classification criteria (Sibbons et al.\ 2015; Menzies et al.\
2015; Boyer et al.\ 2015b). The same selection criteria are also used for classification of x-AGB stars.

The relation between  mass and age for LPVs is presented in the middle panel
of figure~\ref{fig:fig4}, with the coefficients of linear fits listed in
table~\ref{tab:tab6}, for a metallicity of $Z=0.001$.

\begin{table}
\caption[]{Fits to the relation between age and birth mass, $\log t=a\log
M/{\rm M}_\odot+b$.}
\begin{tabular}{ccr}
\hline\hline
\multicolumn{3}{c}{$Z=0.001$} \\
\hline
$a$              & $b$             & validity range                      \\
\hline
$-3.289\pm0.099$ & $9.801\pm0.294$ & $\log{M/{\rm M}_\odot}\leq0.10$      \\
$-2.649\pm0.078$ & $9.743\pm0.291$ & $0.10<\log{M/{\rm M}_\odot}\leq0.31$ \\
$-2.510\pm0.075$ & $9.698\pm0.291$ & $0.31<\log{M/{\rm M}_\odot}\leq0.55$ \\
$-2.132\pm0.063$ & $9.487\pm0.258$ & $0.55<\log{M/{\rm M}_\odot}\leq0.81$ \\
$-1.701\pm0.051$ & $9.137\pm0.273$ & $0.81<\log{M/{\rm M}_\odot}\leq1.11$ \\
$-1.153\pm0.036$ & $8.528\pm0.255$ & $1.11<\log{M/{\rm M}_\odot}\leq1.45$ \\
$-0.678\pm0.020$ & $7.836\pm0.234$ & $\log{M/{\rm M}_\odot}>1.45$         \\
\hline
\end{tabular}
\label{tab:tab6}
\end{table}

We derived the relative pulsation duration (ratio of pulsation timescale to
age) for a given birth mass from the Padova models; these relations are
presented in the right panel of figure~\ref{fig:fig4}, and parameterised by a
set of four Gaussian functions listed in table~\ref{tab:tab7}, for a
metallicity of $Z=0.001$ (relations for other metallicities used in
this paper can be found in the Appendix).

\begin{figure*}
\centerline{\hbox{
\epsfig{figure=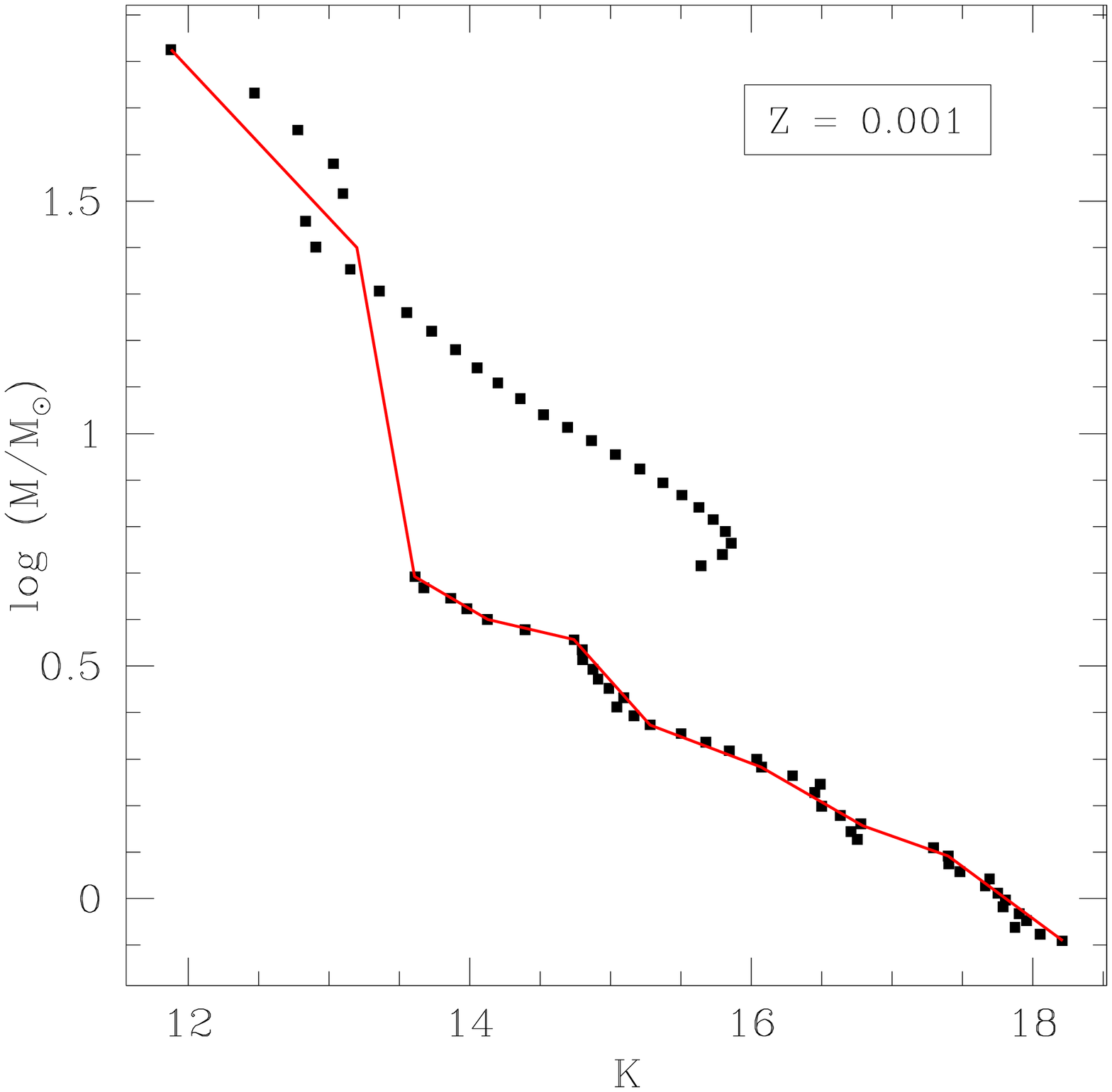,width=58mm}
\epsfig{figure=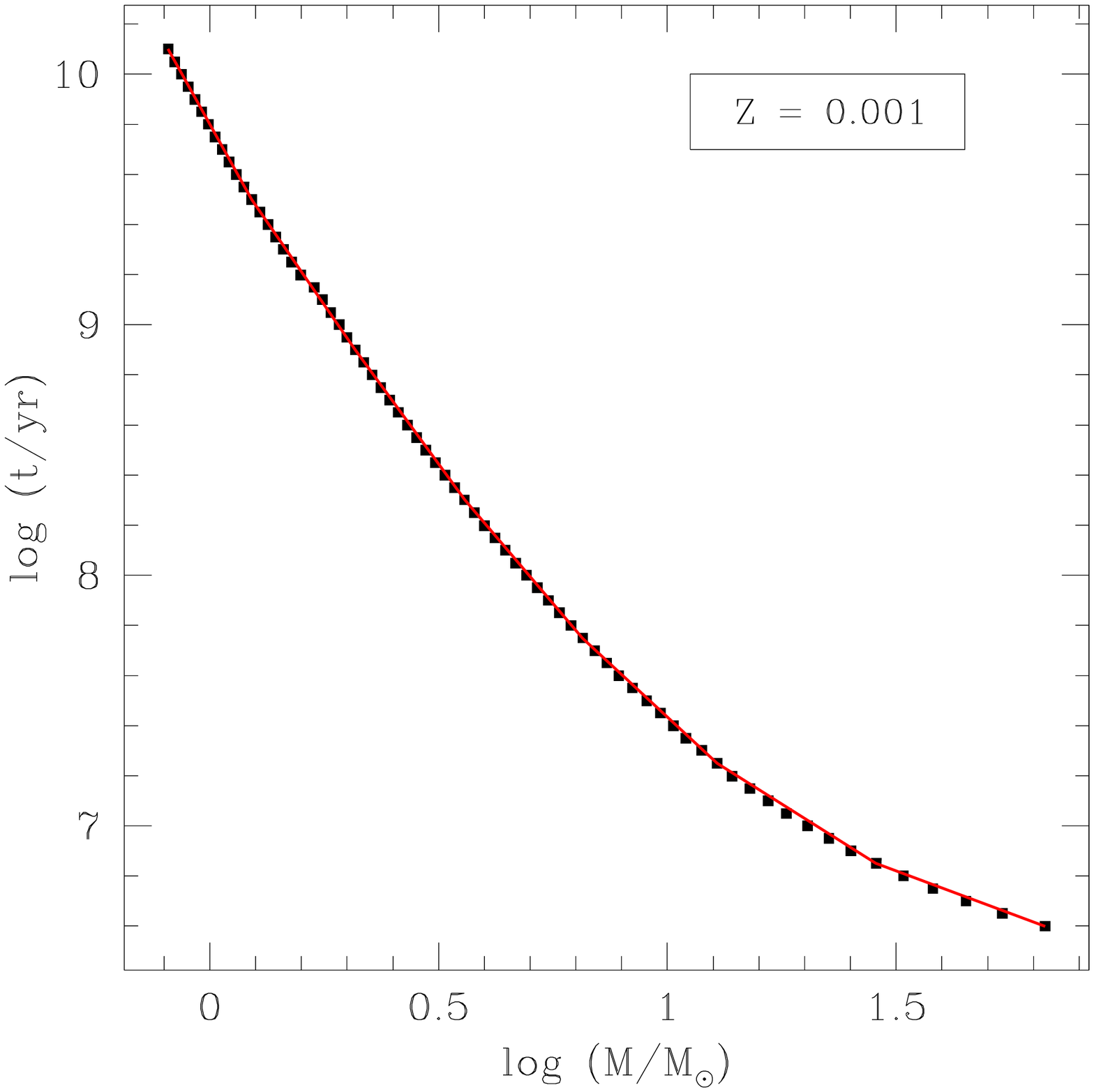,width=58mm}
\epsfig{figure=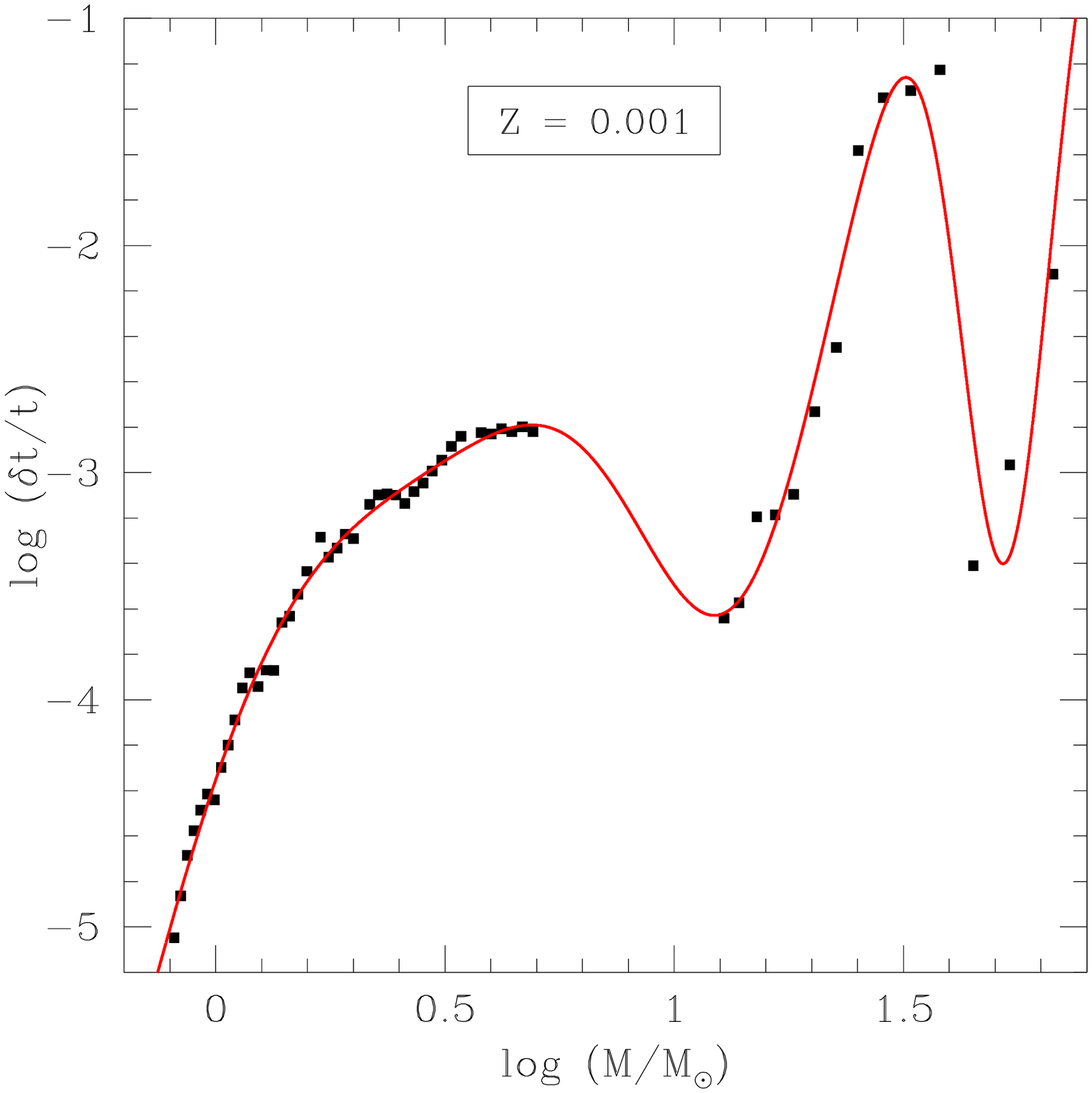,width=58mm}
}}
\caption[]{({\it Left:}) relation between birth mass and K-band
magnitude at the end point of stellar evolution for $Z=0.001$ and a distance
modulus of $\mu=24.37$ mag. Solid lines are fits, in which the function is
interpolated over the super-AGB phase ($0.7\la log{M/{\rm M}_\odot}\la 1.3$).
({\it Middle:}) same as the left, for the relation between age (at the
end point of stellar evolution) and birth mass. ({\it Right:}) same as the
left, for the relation between relative pulsation duration and birth
mass. The points show the ratio of pulsation duration to age, versus mass;
the solid lines are multiple-Gaussian fits, also interpolated over the
super-AGB regime.}
\label{fig:fig4}
\end{figure*}

\begin{table}
\caption[]{Fits to the relation between relative pulsation duration ($\delta
t/t$ where $t$ is the age and $\delta t$ is the pulsation duration) and birth
mass, $\log(\delta t/t)= \Sigma_{i=1}^4a _i\exp\left[-(\log
M/{\rm M}_\odot-b_i)^2/c_i^2\right]$.}
\begin{tabular}{cccc}
\hline\hline
\multicolumn{4}{c}{$Z=0.001$} \\
\hline
$i$ &   $a$    &   $b$    &  $c$  \\
\hline
 1  & $-53.46$ & $-0.115$ & 0.749 \\
 2  & $-02.62$ & $+1.152$ & 0.316 \\
 3  & $+48.48$ & $-0.078$ & 0.697 \\
 4  & $-03.23$ & $+1.722$ & 0.139 \\
\hline
\end{tabular}
\label{tab:tab7}
\end{table}

To calculate the SFH, we thus employ the following procedure:
\begin{itemize}
\item[$\bullet$]{Correct the K-band magnitude for dust attenuation;}
\item[$\bullet$]{Use the corrected K-band magnitude and the
mass--K-band equation (table~\ref{tab:tab5}) to obtain the birth mass;}
\item[$\bullet$]{Use the birth mass and the age--mass equation
(table~\ref{tab:tab6}) to obtain the age;}
\item[$\bullet$]{Use the birth mass and the pulsation duration--mass
equation (table~\ref{tab:tab7}) to obtain the pulsation timescale;}
\item[$\bullet$]{Choose appropriate age bins and apply equation~\ref{eq:eq8}
to calculate the SFR\ in each of those bins.}
\end{itemize}
For each bin, a statistical error can be derived from Poisson statistics as
follows:
\begin{equation}
\sigma_{\xi(t)}=\frac{\sqrt{N}}{N}\xi(t),
\label{eq:eq11}
\end{equation}
where $N$ is the number of stars in each age bin.

\section{Results}
\label{sec:sec4}

Using our sample of evolved stars (section~\ref{sec:sec2-4}) and applying our
method (section~\ref{sec:sec3}), we estimate SFRs in IC\,1613 over the
broad time interval from 30 Myr to $\sim5$ Gyr ago. The observational
reasons for being unable to determine SFRs at earlier epochs will be described
below. Because the metallicity is expected to increase over time as a result
of the chemical evolution driven by nucleosynthesis and mass loss, we first
apply our method assuming different metallicities and then re-analyse
it adopting instead the linear age--metallicity relation (AMR) from Skillman
et al.\ (2014) for values $4\times10^{-4}<Z<4\times10^{-3}$ (corresponding to 13
Gyr ago $<t<$ now).

\begin{figure*}
\centerline{\hbox{
\epsfig{figure=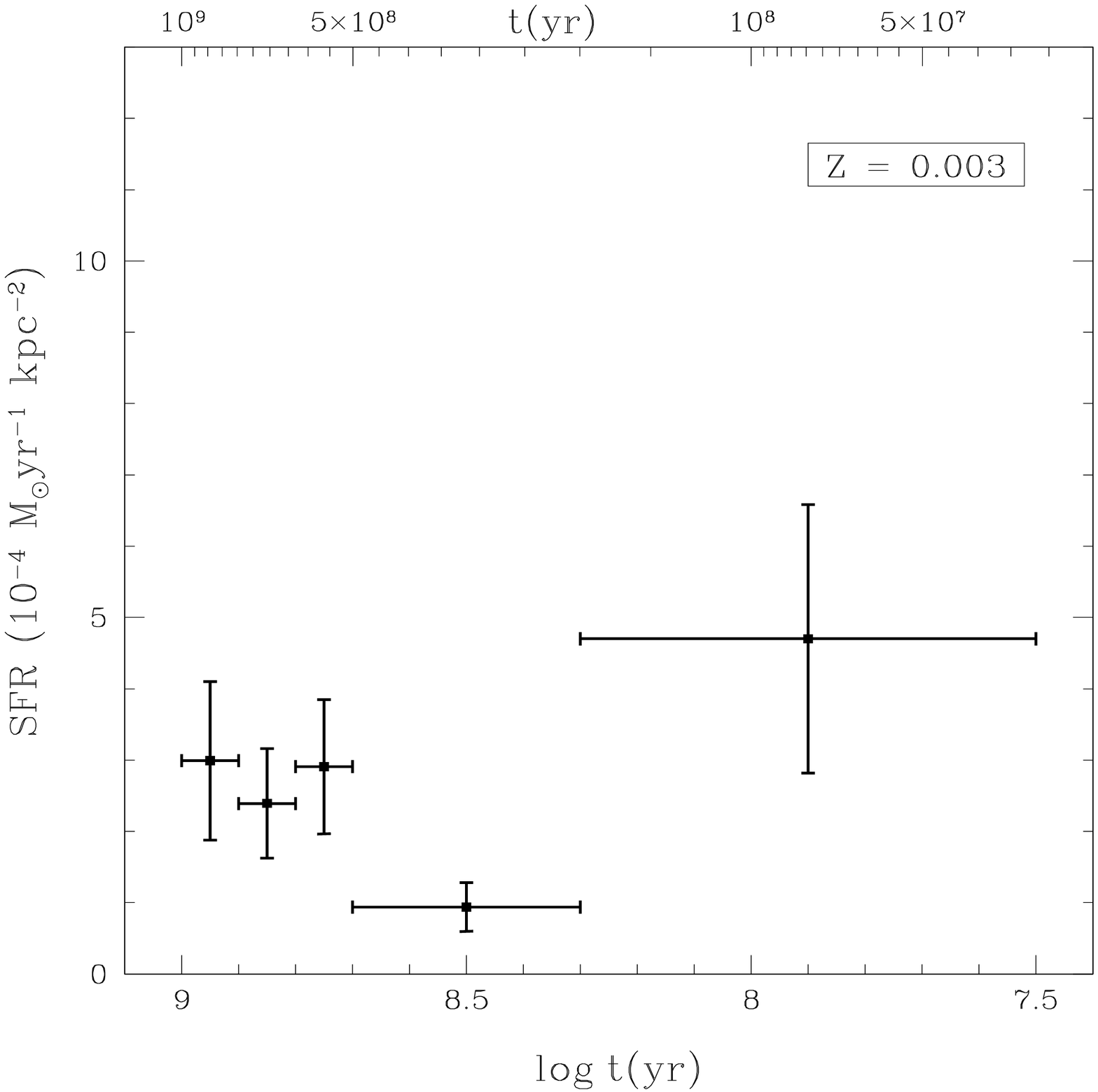,width=88mm}
\epsfig{figure=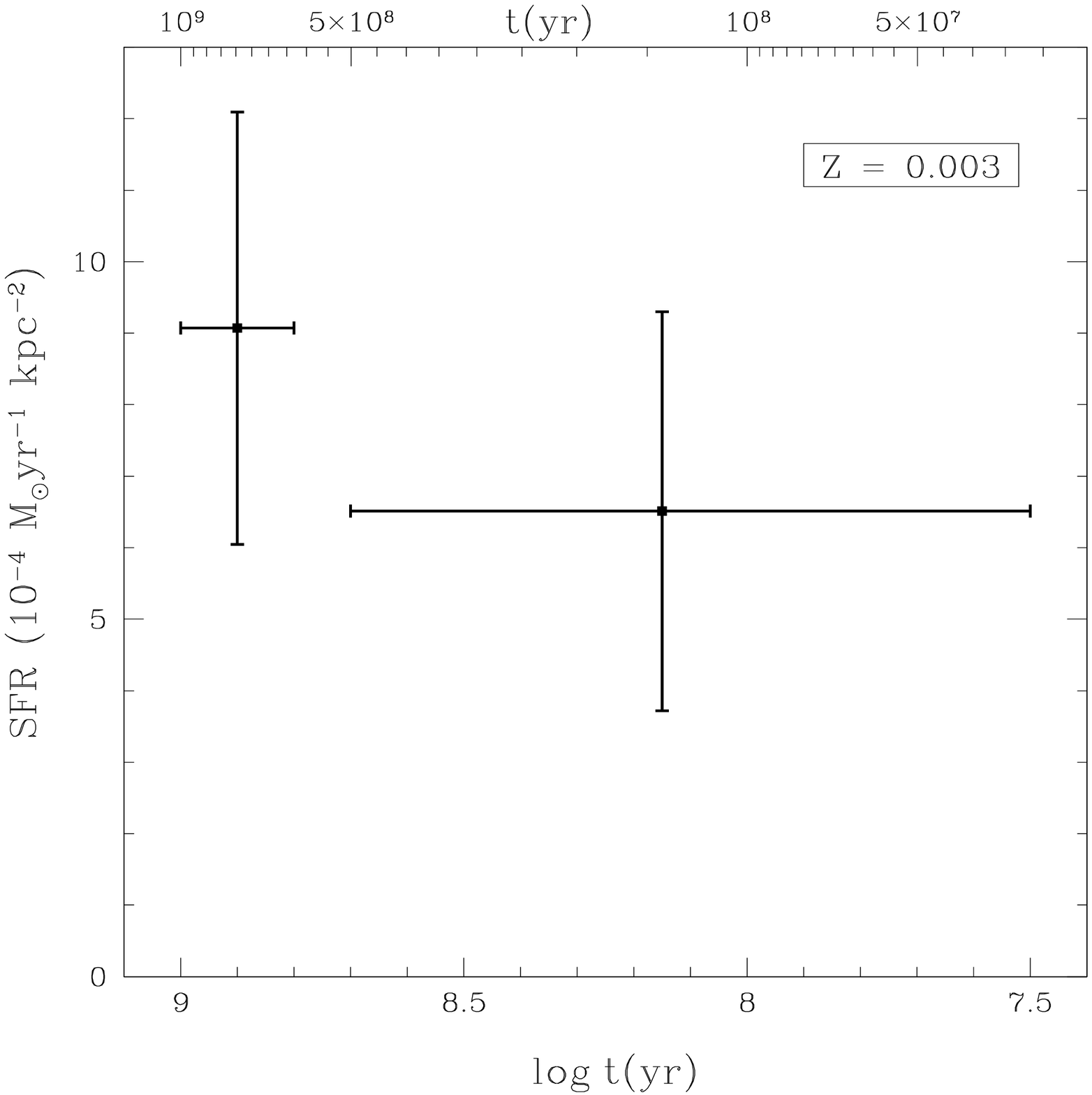,width=88mm}
}}
\caption[]{({\it Left:}) recent SFH (last Gyr) in a $\sim200$ square
arcmin ($\sim11.8$ kpc$^2$) of IC\,1613 assuming a constant metallicity of
$Z=0.003$. ({\it Right:}) same, but in $\sim36$ square arcmin ($\sim2.2$
kpc$^2$) in the central part of the galaxy.}
\label{fig:fig5}
\end{figure*}

We start by examining the recent SFH, at a constant metallicity. The
SFR as a function of look-back time in the last Gyr is shown in the
left panel of figure~\ref{fig:fig5}. For this epoch we assumed $Z=0.003$. The
horizontal bars represent the  spread in age within each bin, while the
vertical bars represent the statistical errors. The mean SFR over the last 200
Myr ($\log t({\rm yr})\sim8.3$) in the central 11.8 kpc$^2$ is
$\xi\sim(0.5\pm0.2)\times10^{-3}$ M$_\odot$ yr$^{-1}$ kpc$^{-2}$;
 it is marginally higher in the central 2.2 kpc$^2$,
$\xi\sim(0.65\pm0.3)\times10^{-3}$ M$_\odot$ yr$^{-1}$ kpc$^{-2}$. This agrees
well with the result derived by Bernard et al.\ (2007): they found a mean SFR
of $\xi\sim(1.6\pm0.8)\times10^{-3}$ M$_\odot$ yr$^{-1}$ kpc$^{-2}$ in the central
part ($r\leq2\rlap{.}^\prime5$) of the galaxy for the last 300 Myr ($\log
t({\rm yr})\sim8.5$), replicating the result obtained by Cole et al.\ (1999)
for a central $\sim0.22$ kpc$^2$ region on the basis of the
main-sequence luminosity distribution. Cole et al.\ also found that the SFR
was $\sim50$\% higher 400--900 Myr ago ($8.6<\log t({\rm yr})<8.9$). While
based on just 19 stars in the central 2.2 kpc$^2$, we find
$\xi\sim(0.9\pm0.3)\times10^{-3}$ M$_\odot$ yr$^{-1}$ kpc$^{-2}$ 500--900 Myr ago
(Fig.~\ref{fig:fig5}, right), i.e.\  a very similar SFH as that
determined by Cole et al.

\begin{figure}
\centerline{\epsfig{figure=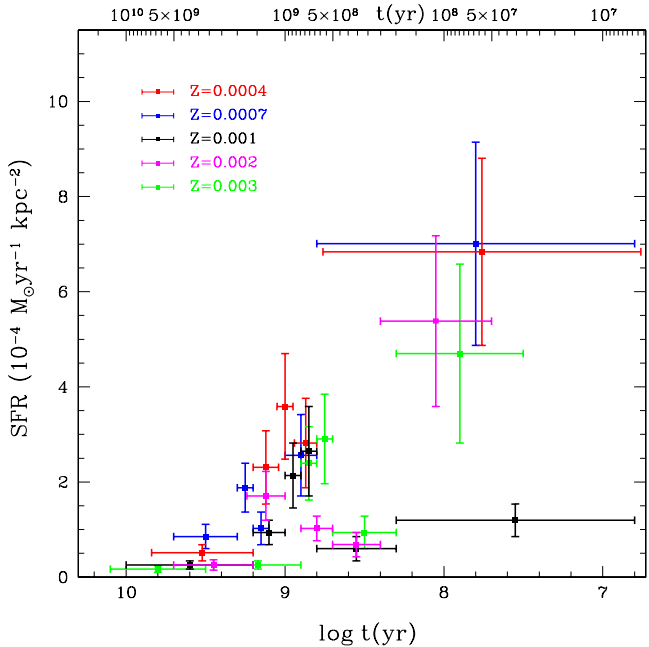,width=85mm}}
\caption[]{IC\,1613 SFHs for different metallicities.}
\label{fig:fig6}
\end{figure}

Subsequently, we applied this process for all epochs and five metallicities
($Z=0.0004$, 0.0007, 0.001, 0.002, 0.003); the result is shown in
figure~\ref{fig:fig6}. Clearly, the result is sensitive to the adopted
metallicity, so in order to trace back the SFH we need to take into account
the variation of metallicity with look-back time.

\begin{figure}
\centerline{\epsfig{figure=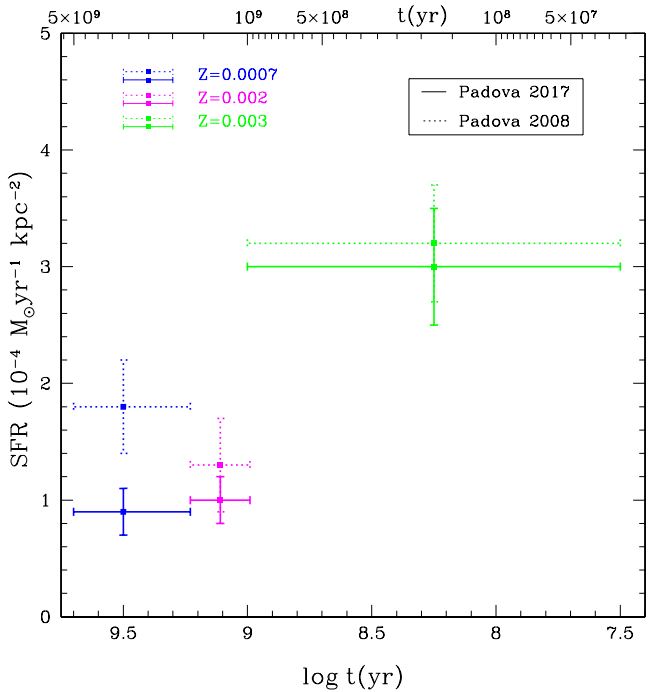,width=85mm}}
\caption[]{The SFH in IC\,1613 with the adoption of an AMR akin that derived
by Skillman et al.\ (2014), using the pulsation durations from Marigo et al.\
(2017) (solid) and Marigo et al.\ (2008) (dotted).}
\label{fig:fig7}
\end{figure}

Skillman et al.\ (2014) modelled the CMDs in a small field near the half-light
radius to determine the SFH for the full 13 Gyr of evolution. They found that
most -- if not all -- stars in IC\,1613 formed after the epoch of reionization
($\sim12.8$ Gyr ago), without a dominant formation epoch. To compare with
their result, we adopt a similar AMR as they did: $Z=0.003$ for the last Gyr
($\log t({\rm yr})<9$), $Z=0.002$ for 1 Gyr $<t< 2$ Gyr ($9<\log t({\rm
yr})<9.3$) and $Z=0.0007$ for $t>2$ Gyr ($\log t({\rm yr})>9.3$). We thus find
a mean value of the SFR across IC\,1613 over the last Gyr of
$\xi=(3.0\pm0.5)\times10^{-4}$ M$_\odot$ yr$^{-1}$ kpc$^{-2}$
(Fig.~\ref{fig:fig7}), in excellent agreement with Skillman et al., who found
$\xi\sim3.4\times10^{-4}$ M$_\odot$ yr$^{-1}$ kpc$^{-2}$.

However, Skillman et al.\ (2014) found SFRs at earlier epochs (1 Gyr $<t<2$
Gyr and 2 Gyr $<t<5$ Gyr) that are two to three times higher than our results,
and we could not trace star formation beyond $\sim5$ Gyr ago. We now
discuss how the SFH we derive for look-back times $t>9$ Gyr may have been
biased.

The truncated SFH (beyond 5 Gyr) is mainly due to observational
limitations with regard to our method:
\begin{itemize}
\item[$\bullet$]{The sample of LPVs comes from the survey by Menzies et al.\
(2015), which quote a completeness limit of $K_{\rm s}\approx18$ mag and will
have missed red variables with $(H-K_{\rm s})>2.0$ mag and $K_{\rm s}>16.3$
mag. In order to trace the SFH to $t>5$ Gyr ($\log t({\rm yr})>9.7$) at low
metallicity ($Z\sim0.0007$), however, we need stars that have
dereddened K-band magnitudes fainter than $\sim18$ mag (see the isochrone
diagram in figure~\ref{fig:fig11}).}
\item[$\bullet$]{Regarding x-AGB stars, the {\it Spitzer} sample from Boyer et
al.\ (2015a) is not flux limited, and the near-IR data from Sibbons et al.\
(2015) are deep enough ($J<19.7$, $H<19.3$ and $K<18.9$ mag) to provide
counterparts to place them in the CMD. However, most x-AGB stars are carbon
stars, with the odd more massive OH/IR star -- cf.\ Woods et al.\ (2011), and
carbon stars arise from stars born not longer than $\sim5.5$ Gyr ago (Dell
Agli et al.\ 2016). Thus, these x-AGB samples do not make up for the
incompleteness of the variability survey by Menzies et al.\ beyond 5 Gyr.}
\end{itemize}

\begin{figure*}
\centerline{\hbox{
\epsfig{figure=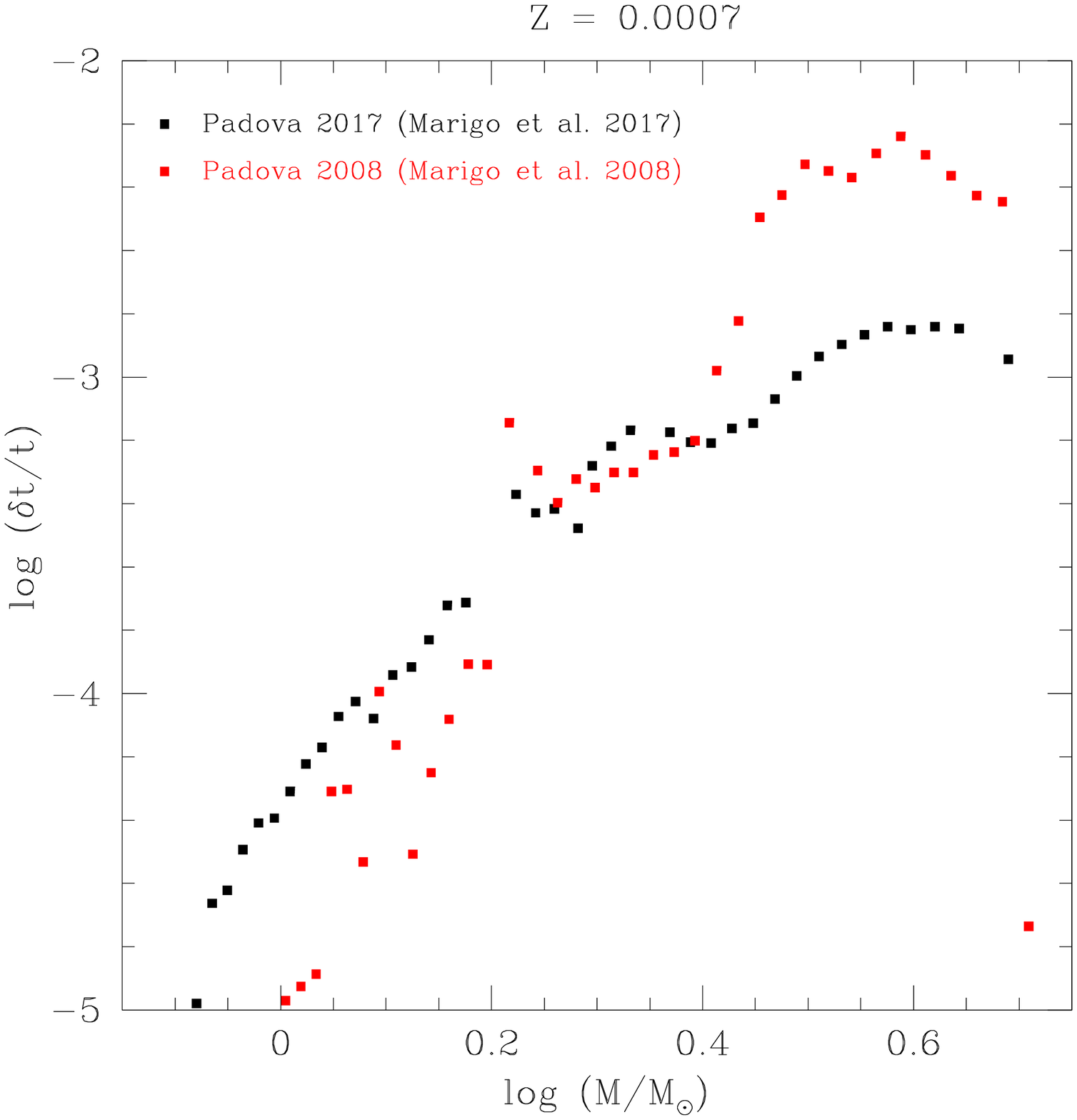,width=88mm}
\epsfig{figure=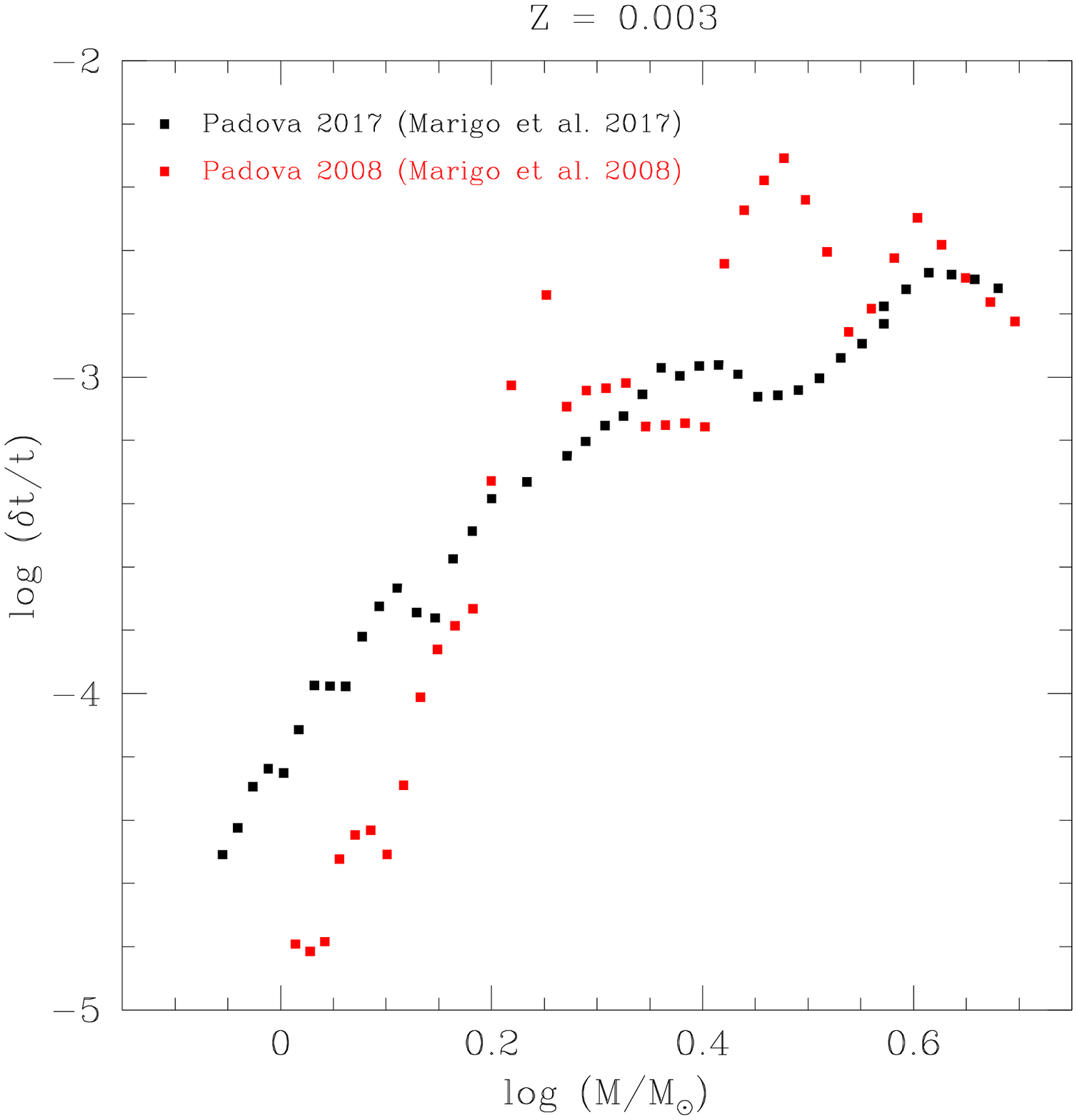,width=88mm}
}}
\caption[]{The differences between the pulsation durations ($\delta t$) from
the most recent Padova models (Marigo et al.\ 2017, black) and previous
version (Marigo et al.\ 2008, red) $Z=0.0007$ ({\it left}) and $Z=0.003$ ({\it
right}).}
\label{fig:fig8}
\end{figure*}

The reasons for the discrepancy between our SFR determinations a few Gyr ago
and those by Skillman et al.\ (2014) are  more subtle, and probably include
the following:
\begin{itemize}
\item[$\bullet$]{It is likely that the completeness limits of the monitoring
survey and possibly the near-IR complement to the {\it Spitzer} survey start
to deplete our sample of sources at multi-Gyr ages. This is not an issue for
the most recent Gyr, which is confirmed by the good correspondence
with the results in the literature as described above.}
\item[$\bullet$]{While the field studied by Skillman et al.\ (2014) may be
representative of the bulk of IC\,1613 in recent times, this may not be true
at earlier epochs. We tried to find the SFR on the same location as their
field ($RA\sim1^{\rm h}4^{\rm m}30^{\rm s}$ and $Dec\sim2^\circ9^\prime$), but the
sample became limited to just one star and this did not yield a meaningful
SFR.}
\item[$\bullet$]{The pulsation durations ($\delta t$) may have been
overestimated, which would immediately result in an underestimated SFR.
Figure~\ref{fig:fig8} shows the differences between the Marigo et al.\ (2017)
and Marigo et al.\ (2008) pulsation durations. For stars with $0.3 \la
\log M/{\rm M}_\odot\la 0.7$ the pulsation duration was shortened, while for
intermediate- and low-mass stars with $\log M/{\rm M}_\odot \la 0.3$ the
duration was lengthened. Our results are in better agreement with the results
from Skillman et al.\ (2014) when using the 2008 models (dotted lines in
figure~\ref{fig:fig7}). For $t>2$ Gyr ($\log t({\rm yr})>9.3$) the SFR using
the Marigo et al.\ (2008) pulsation durations doubles and becomes$\sim60$\% of
the SFR derived by Skillman et al.\ ($\xi\sim3\times10^{-4}$ M$_\odot$ yr$^{-1}$
kpc$^{-2}$); for 1 Gyr $<t<2$ Gyr ($9<\log t({\rm yr})<9.3$) the SFR would
also increase but only to $\sim30$\% of the SFR derived by Skillman et al.\
($\xi\sim3.6\times10^{-4}$ M$_\odot$ yr$^{-1}$ kpc$^{-2}$). This is not the first
time that results suggest that the pulsation durations in the Padova models
may be too generous: Javadi et al.\ (2017) (see also Javadi et al.\ 2013)
found corrections of a factor $\sim3$ were needed at solar and slightly
sub-solar metallicity. It appears that the 2008 Padova models are preferred
over the 2017 models, at least for older ages at low metallicities, but still
need to be corrected by a factor $\sim2$--3.}
\end{itemize}

In any case we do not find a discrete epoch of enhanced (or suppressed) star
formation (Fig.~\ref{fig:fig7}), that could be linked to external
triggering mechanisms. This supports the notion that IC\,1613 has evolved in
isolation for at least the past 5 Gyr (Skillman et al.\ 2014; Cole et
al.\ 1999; Stinson et al.\ 2007).

\section{Discussion}
\label{sec:sec5}

\subsection{Galactocentric radial gradient of the SFH}
\label{sec:sec5-1}

Galactocentric radial gradients are imprinted with the dynamical history and
propagation of star formation. It appears that stellar population gradients
are universal in dwarf galaxies, in the sense that the mean age of the stellar
population is younger towards the centre of the galaxy (e.g., Skillman et al.\
2014; Hidalgo et al.\ 2013).

\begin{figure}
\centerline{\epsfig{figure=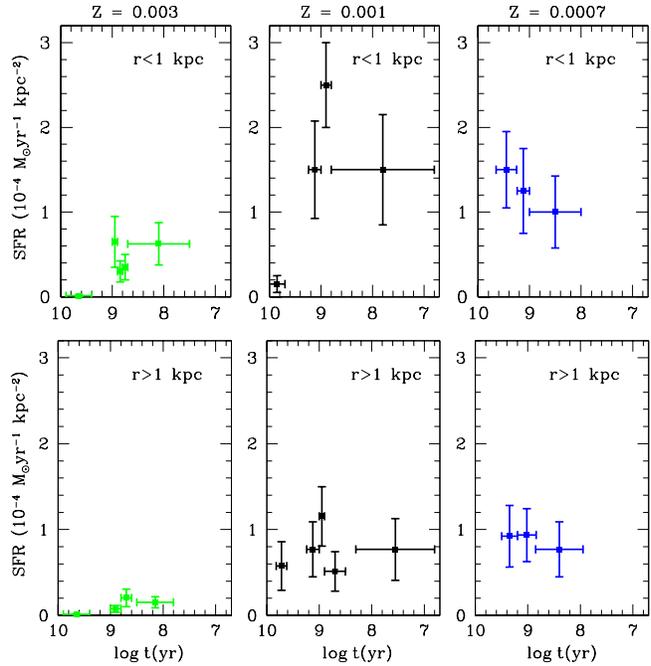,width=85mm}}
\caption[]{SFHs of two regions at different galactocentric radii ($r<1$ kpc,
$r>1$ kpc), for different metallicities.}
\label{fig:fig9}
\end{figure}

To examine radial gradients in IC\,1613, we divided our sample into two parts
-- an inner ($r<1$ kpc) and an outer ($r>1$ kpc) region. The resulting SFHs
are shown in figure~\ref{fig:fig9}. In the inner part, the mean SFR over the
past Gyr ($Z\sim 0.003$) was $\xi\sim0.7\times10^{-4}$ M$_\odot$ yr$^{-1}$
kpc$^{-2}$, in the outer part it was $\xi\sim0.2\times10^{-4}$ M$_\odot$
yr$^{-1}$ kpc$^{-2}$. Clearly, the inner part contains more young stars
relative to the outer region, but this is in part due to the radially
decreasing overall stellar density. However, when considering the SFRs at
older times and lower metallicity, they increase more rapidly in the outer
parts than in the inner parts of IC\,1613. In fact, at the lower metallicities
($Z\sim0.0007$) the SFR in the outer galaxy rivals that near the centre.

Another way of quantifying this gradient is in terms of the ratio of SFRs at
older times ($\log t({\rm yr})>9.2$, $Z=0.0007$) to those at recent times
($\log t({\rm yr})<9$, $Z=0.003$). In the inner region this ratio is $\sim2$,
while in the outer region it is $\sim7$. Considering the errorbars
(minimum of old SFR and maximum of recent SFR) decreases the ratio to $\sim1$
for the inner part and $\sim3$ for the outer part. This outside--in evolution
scenario (at least for the most recent $\sim5$ Gyr) is in agreement with what
is typically found in dwarf galaxies (e.g., Hidalgo et al.\ 2003,
2008, 2013) and differs from what is found in the low-mass spiral galaxy
M\,33, for instance (Javadi et al.\ 2017).

As we described in the result section, because of incompleteness of
the surveys it is possible that we have underestimated the SFR older than a
Gyr. Likewise, the discrepant pulsation durations between the 2008 and 2017
Padova models could have affected the older SFRs. But this is true for both
the inner and outer part in this section, and we do not expect the observed
gradient to change. Furthermore, the SFH over the last Gyr in the central part
(right panel of Fig.~\ref{fig:fig5}) compared to that in our larger field of
view (left panel of Fig.~\ref{fig:fig5}) shows an obvious concentration of
recent star formation in the central part of the galaxy.

\subsection{The origin of the main hydrogen supershell}
\label{sec:sec5-2}

The neutral interstellar medium within dwarf galaxies often shows holes, arcs,
and shells. These structures are typically explained by the (combined) effects
of stellar winds from massive stars and supernova (SN) explosions
(Dyson \& de Vries 1972; Weaver et al.\ 1977), but this scenario
struggles to explain the largest structures (Silich et al.\ 2006 and
references therein).

With regard to IC\,1613, Silich et al.\ (2006) identified several H\,{\sc i}
supershells, including ``main supershell'' -- an H\,{\sc i} hole of 1
kpc in diameter and surrounded by an H\,{\sc i} ring centred at
$RA=1^{\rm h}4^{\rm m}52^{\rm s}$ ($16\rlap{.}^\circ2167$), $Dec=2^\circ7^\prime$
($2\rlap{.}^\circ1167$) (see Fig. \ref{fig:fig3}). The H\,{\sc i} mass of this
supershell is $2.8\times10^7$ M$_\odot$. The absence of regular expansion and
the thickness of the ring led Silich et al.\ to believe the shell has stalled.
The wide range in age of the OB associations -- up to 30 Myr (Borissova et
al.\ 2004) -- found inside the shell, suggested to them that the shell was
formed over a period of $\sim30$ Myr.

Silich et al.\ simulated the formation of the main supershell for two
different (thin and thick) galactic disc models, in both cases concluding a
formation timescale of 30 Myr. They thus estimated a mid-plane gas number
density $n_0\sim2.8(1.4)$ cm$^{-3}$ and a SFR of $xi\sim7.5(2.3)\times10^{-3}$
M$_\odot$ yr$^{-1}$ for their thin (thick) disc models. This is an order of
magnitude larger than the in-situ SFR they derived from H$\alpha$ data,
$xi\sim(3$--$4)\times10^{-4}$ M$_\odot$ yr$^{-1}$, leading them to reject the
multiple-SN scenario.

\begin{figure}
\centerline{\epsfig{figure=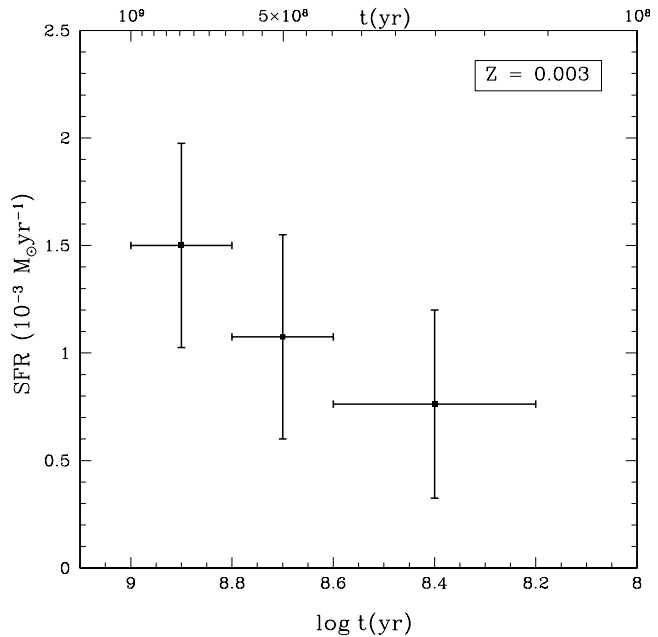,width=85mm}}
\caption[]{SFHs within the main supershell (Silich et al.\ 2006) of IC\,1613
assuming a constant metallicity of $Z=0.003$.}
\label{fig:fig10}
\end{figure}

To revisit the SFR within the main supershell, we selected stars -- 15 in
total -- from our sample that reside inside the contour outlining the main
supershell in Silich et al.\ (2006) (Fig.~\ref{fig:fig3}). Assuming a
constant metallicity of $Z=0.003$, we thus derive the recent SFH within the
shell (Fig.~\ref{fig:fig10}). Because of the small number statistics we lack
supergiants within this limited sample and thus cannot ascertain the SFR at
$t<100$ Myr. However, the SFR seems to have slowly but steadily decreased in
recent times, and if this trend continued into the most recent 30 Myr we would
estimate a SFR that is compatible with the one derived from H$\alpha$
by Silich et al.\ (2006). One may wonder, though whether the shell could in
fact have formed over much longer timescales, in excess of 100 Myr -- from
which OB associations no longer exist.

\subsection{Ability of mass loss to sustain star formation}
\label{sec:sec5-3}

Star formation depletes a galaxy of its gas, but stellar mass loss replenishes
it. The question therefore is whether the latter can sustain the former.

Silich et al.\ (2006) derived a total H\,{\sc i} mass in IC\,1613 of
$6.0\times10^7$ M$_\odot$, in good agreement with previous estimates
 from Lake \& Skillman (1989). Considering also the contribution from
helium, the total mass of the ISM would be $\sim9.0\times10^{7}$ M$_\odot$.
Dell'Agli et al.\ (2016) used stellar evolution models to estimate the dust
production rate by AGB stars within IC\,1613, $\sim6\times10^{-7}$ M$_\odot$
yr$^{-1}$, and gas-to-dust mass ratio, $\sim1000$, and hence a total mass
return rate of $6\times10^{-4}$ M$_\odot$ yr$^{-1}$.

The recent SFR of $\xi=(5.5\pm2)\times10^{-3}$ M$_\odot$ yr$^{-1}$ we derived
from the evolved star population (Fig.~\ref{fig:fig5}) exceeds the mass return
rate by an order of magnitude. This means that the ISM will be depleted by
star formation on a timescale of 16.4 Gyr. Accounting for the mass return rate
would stretch this by a small amount to $\sim18.4$ Gyr. Considering
the contribution of ionized mass of the galaxy (about a few times $10^6$ from
Silich et al.\ 2006) can increase the estimation up to 10\%.

This assumes that star formation is 100\% efficient, and that no ISM gas is
blown out of the shallow gravitational potential well of the dwarf galaxy, so
star formation at the current rate would unlikely continue for that long. On
the other hand, the above estimate also neglects contributions to mass return
by SNe, luminous blue variables et cetera -- though these do not make a huge
difference when assessed over cosmological times (cf.\ Javadi et al.\ 2013).
The expectation therefore is that IC\,1613 cannot continue to form stars at
the current rate for more than a few Gyr, unless gas is accreted from outside,
for instance when (if) it traverses the halo from M\,31 or the Milky Way, or
intergalactic gas if such reservoirs exist.

\section{Summary of conclusions}
\label{sec:sec6}

Selecting stars at the end points of their evolution, we applied a recently
developed method to derive the SFH in the isolated Local Group dIrr galaxy
IC\,1613. Our main findings are:
\begin{itemize}
\item[$\bullet$]{From a combination of near-/mid-IR CMDs we identified
21 x-AGB stars, of which 2 stars had not been identified before.}
\item[$\bullet$]{We do not find any dominant period of star formation over the
past 5 Gyr, which suggests that IC\,1613 may have evolved in isolation for at
least that long.}
\item[$\bullet$]{The SFH over the past few Gyr confirms those derived by other
methods, with a radial gradient that indicates the mean age of the central
population is younger than that in the outskirts.}
\item[$\bullet$]{The extrapolated rate of recent star formation within
the main H\,{\sc i} supershell falls short by an order of magnitude of the
required recent SFR to have produced the shell.}
\item[$\bullet$]{The current reservoir of interstellar gas may sustain star
formation at the current rate for several Gyr into the future; however the
rate at which mass is returned will not extend that time for very much longer
-- indeed, star formation beyond the next few Gyr will diminish and eventually
cease altogether unless gas can be accreted from outside.}
\end{itemize}

\section*{Acknowledgments}

We are deeply grateful to Sohrab Rahvar and Martha Boyer for their
valuable comments, and discussions. We thank Mehrdad Phoroutan mehr for his
kind comments. We would also like to thank Sergiy Silich, Tatyana Lozinskaya
and Alexei Moiseev for sharing their H\,{\sc i} maps and their responses to
our questions. Finally, The authors are indebted to the anonymous referee for
the careful reading of the manuscript and for the helpful comments, which
prompted us to improve the quality of this work.

\appendix
\section{Supplementary material}
\label{app:app}

In this section we present figures and fits to the K-band--mass, age--mass and
pulsation duration--mass relation, derived from the Padova models (Marigo et
al.\ 2017; Marigo et al.\ 2008). We used these equations in our analysis to
find the SFH of IC\,1613. For more detail see section~\ref{sec:sec3} where we
explain the method and present the figures and fits for the case of $Z=0.001$
(and $\mu=24.37$ mag). Here we also present isochrones for $Z=0.0007$.

\begin{figure*}
\centerline{\hbox{
\epsfig{figure=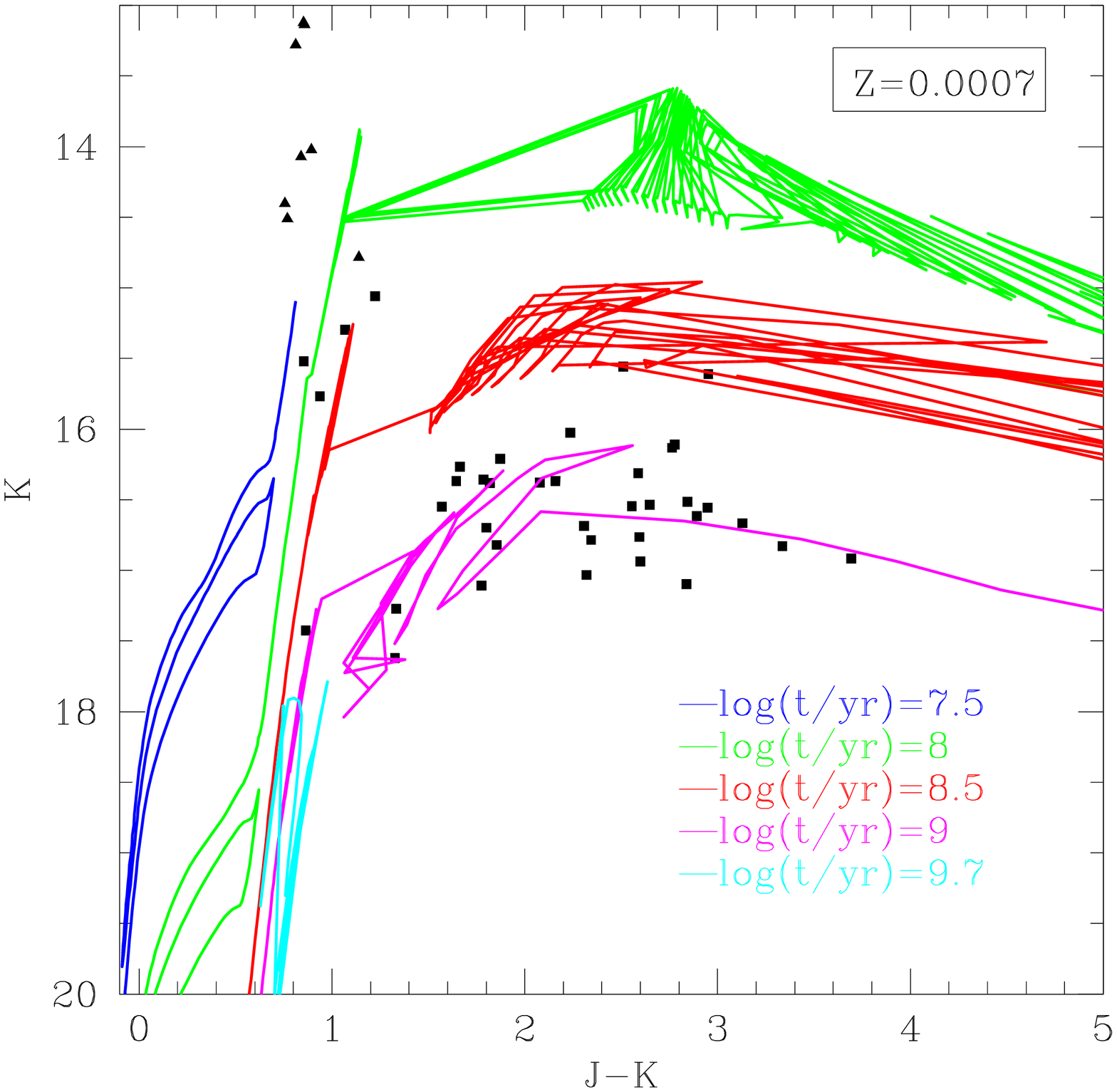,width=88mm}
\epsfig{figure=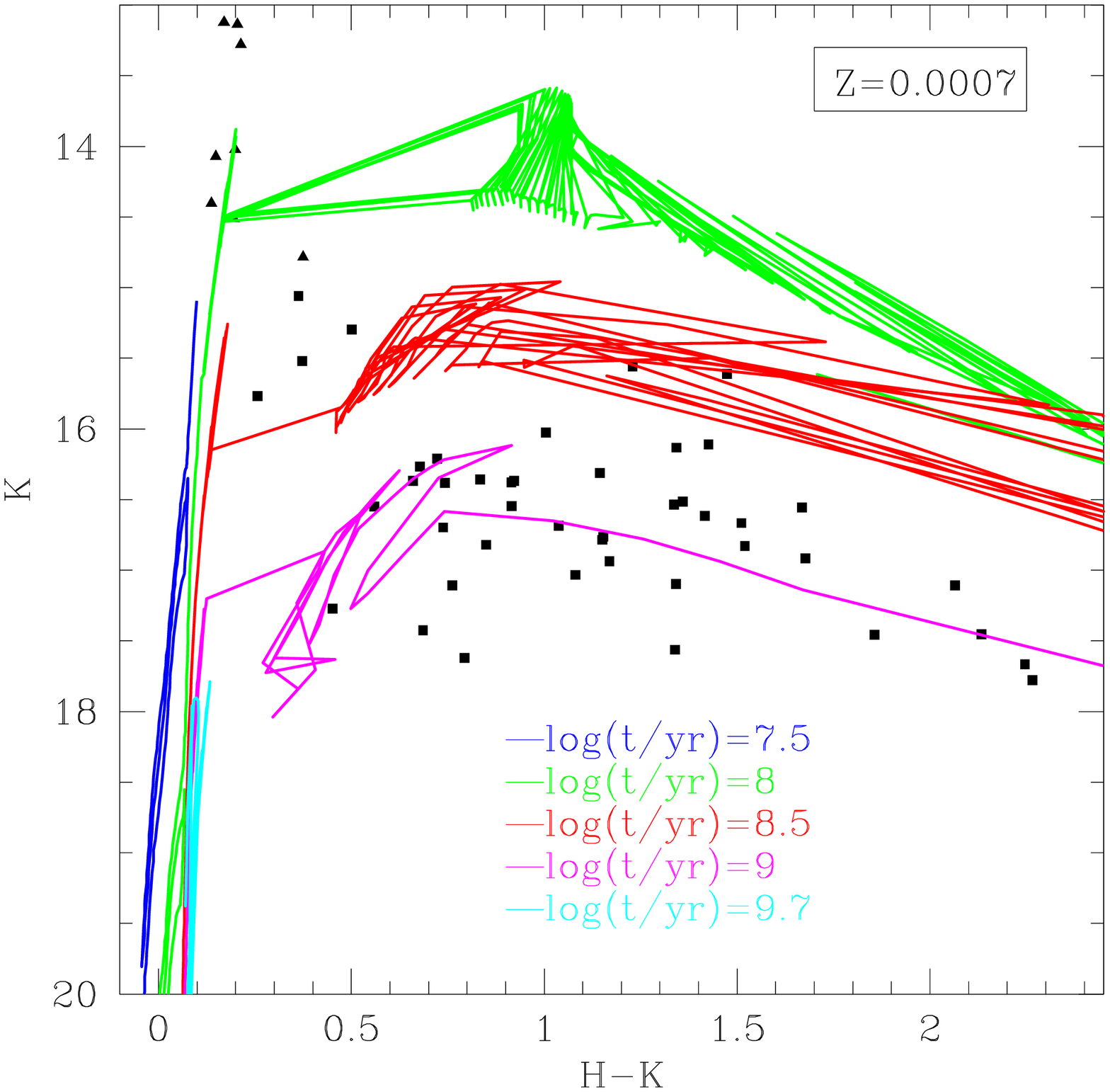,width=88mm}
}}
\caption[]{Same as figure~\ref{fig:fig2} but for $Z=0.0007$.}
\label{fig:fig11}
\end{figure*}

\begin{figure*}
\centerline{\hbox{
\epsfig{figure=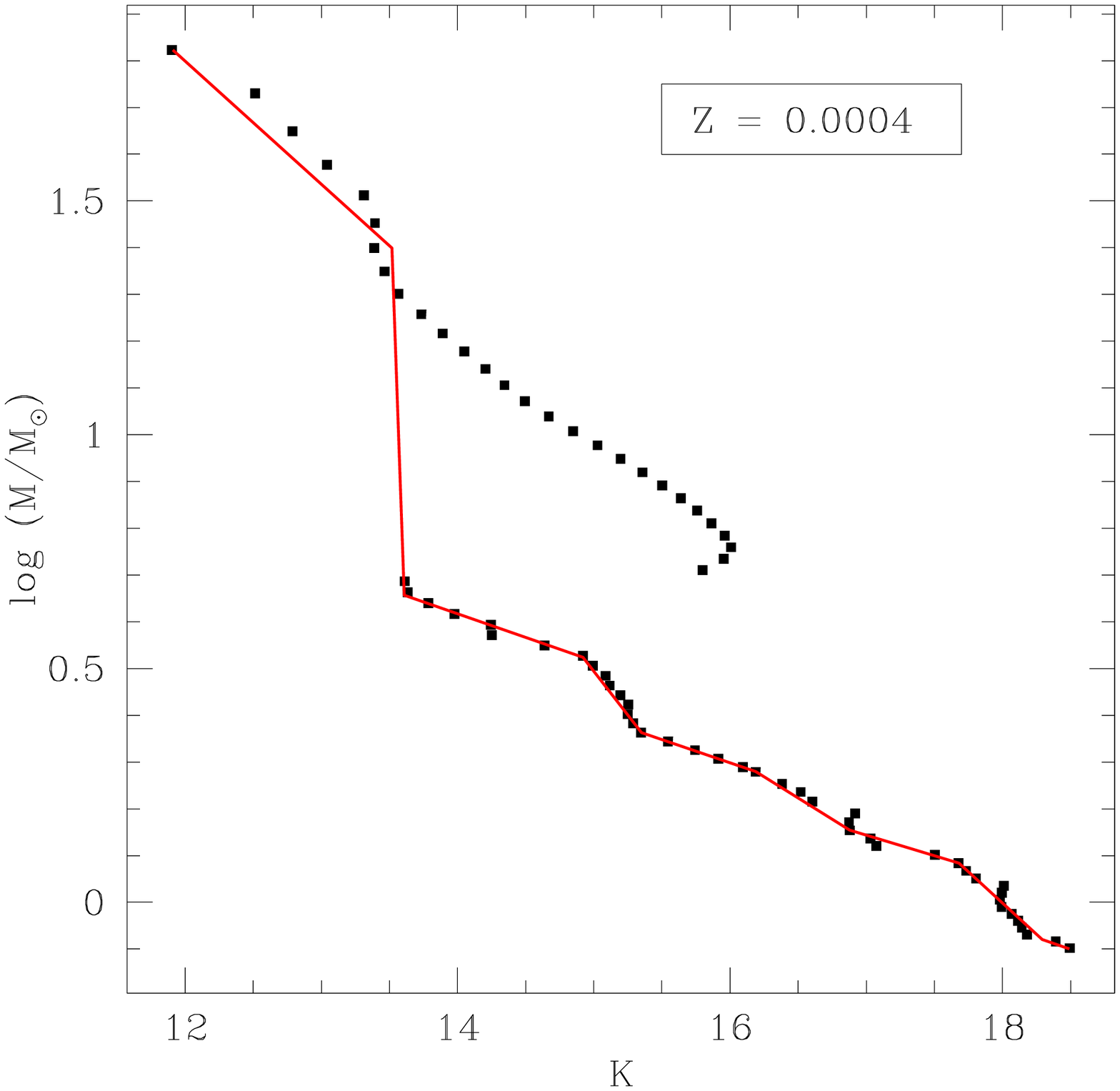,width=58mm}
\epsfig{figure=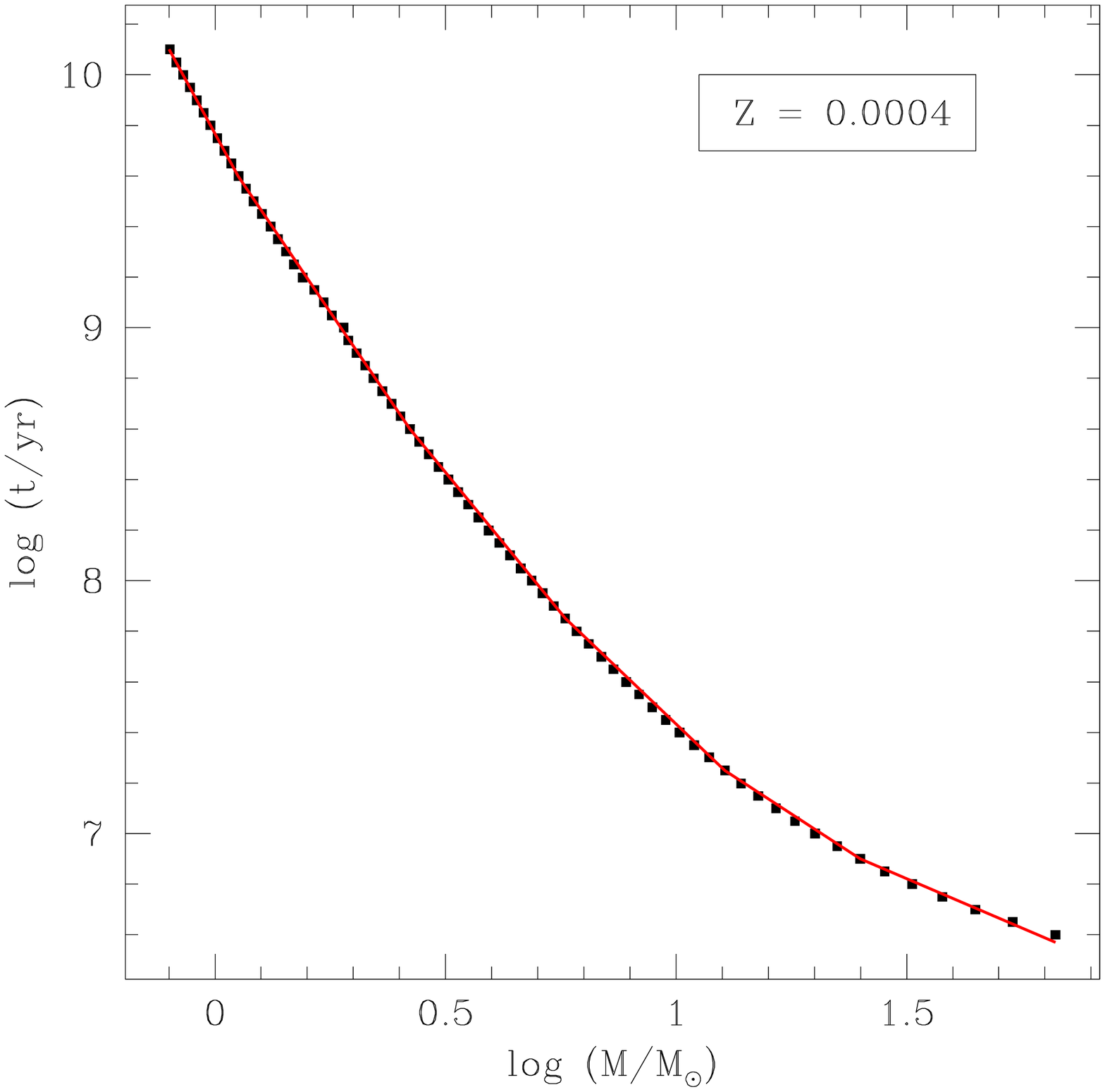,width=58mm}
\epsfig{figure=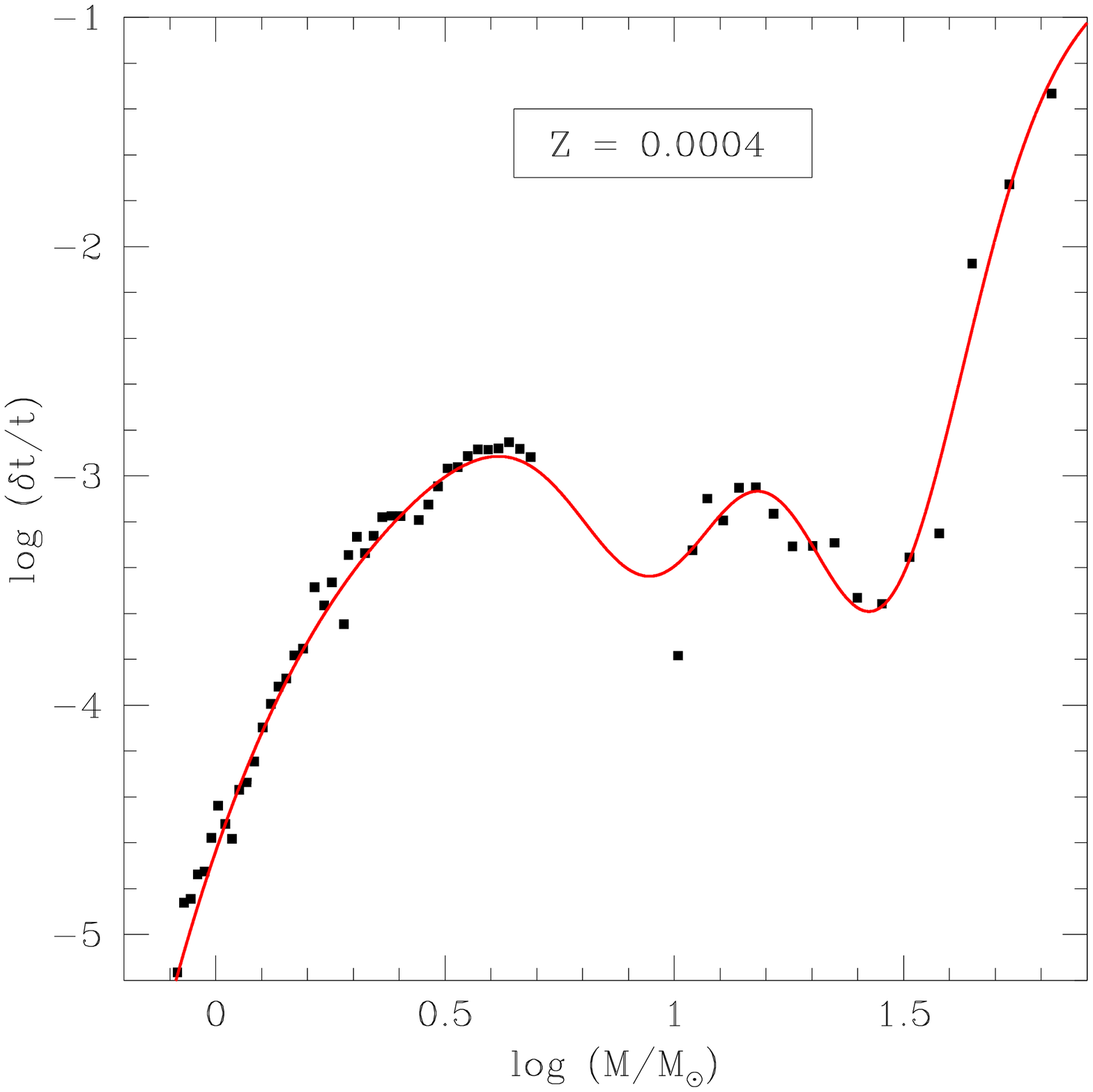,width=58mm}
}}
\caption[]{Same as figure~\ref{fig:fig4} but for $Z=0.0004$.}
\label{fig:fig12}
\end{figure*}

\begin{figure*}
\centerline{\hbox{
\epsfig{figure=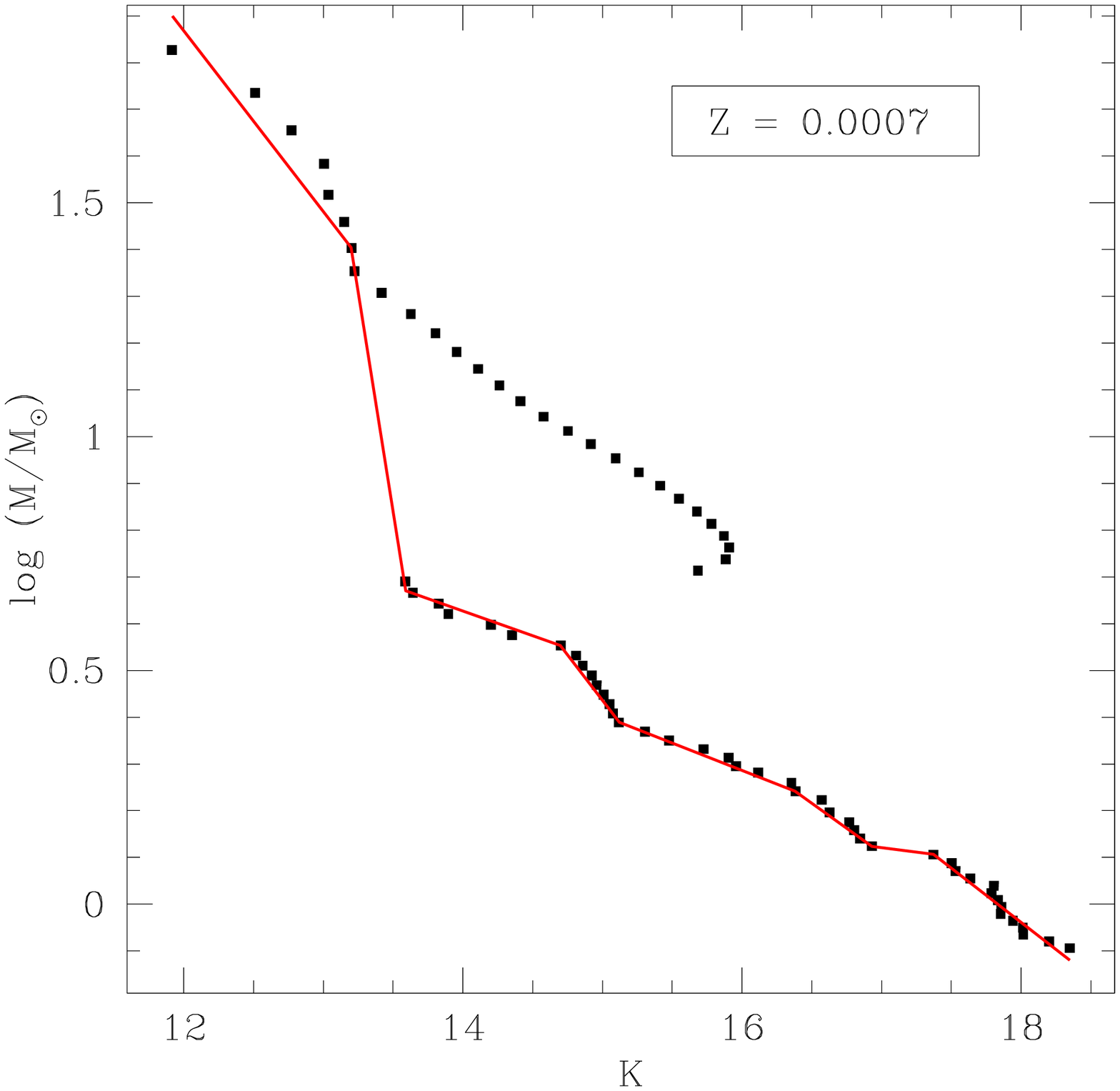,width=58mm}
\epsfig{figure=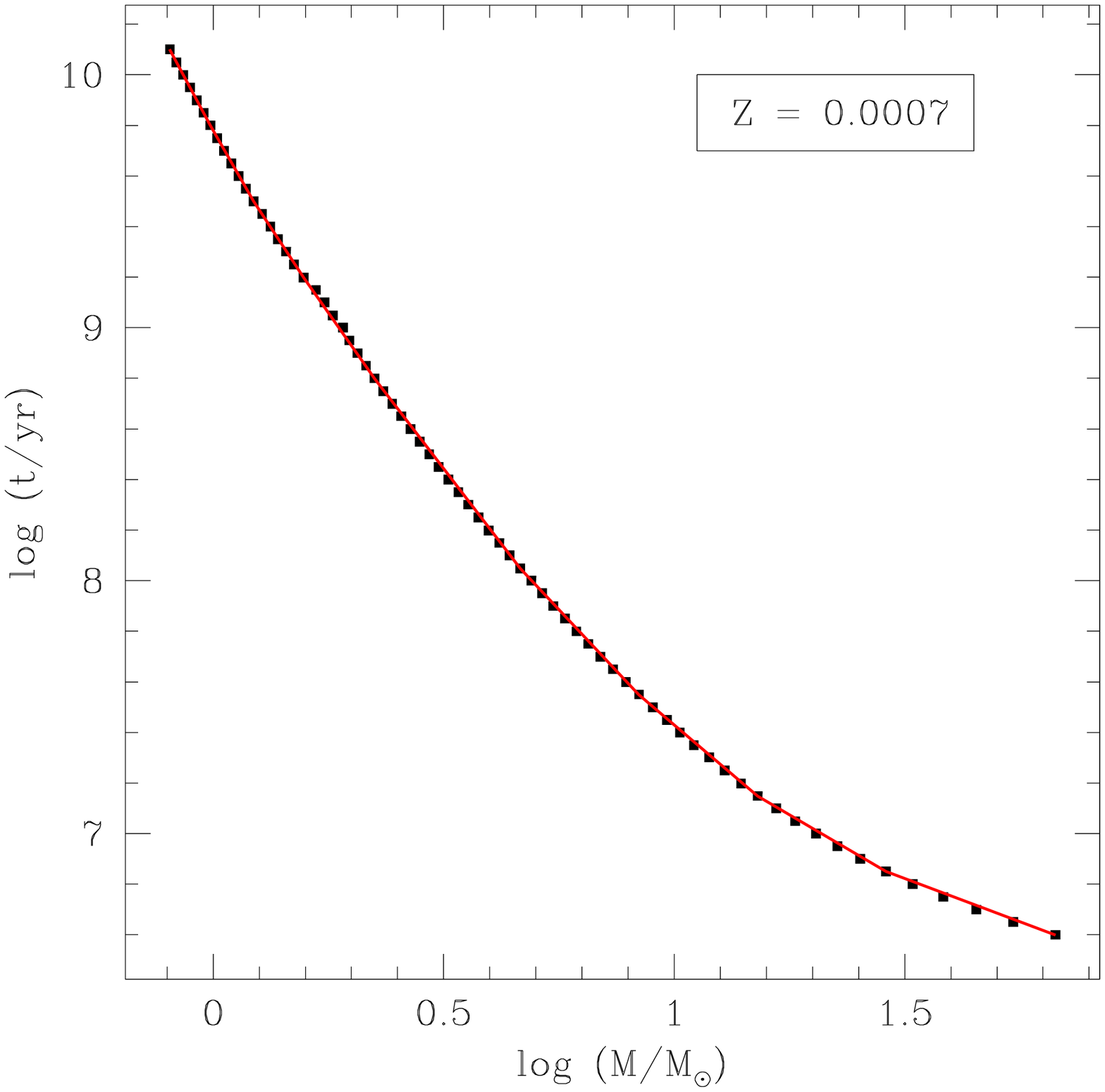,width=58mm}
\epsfig{figure=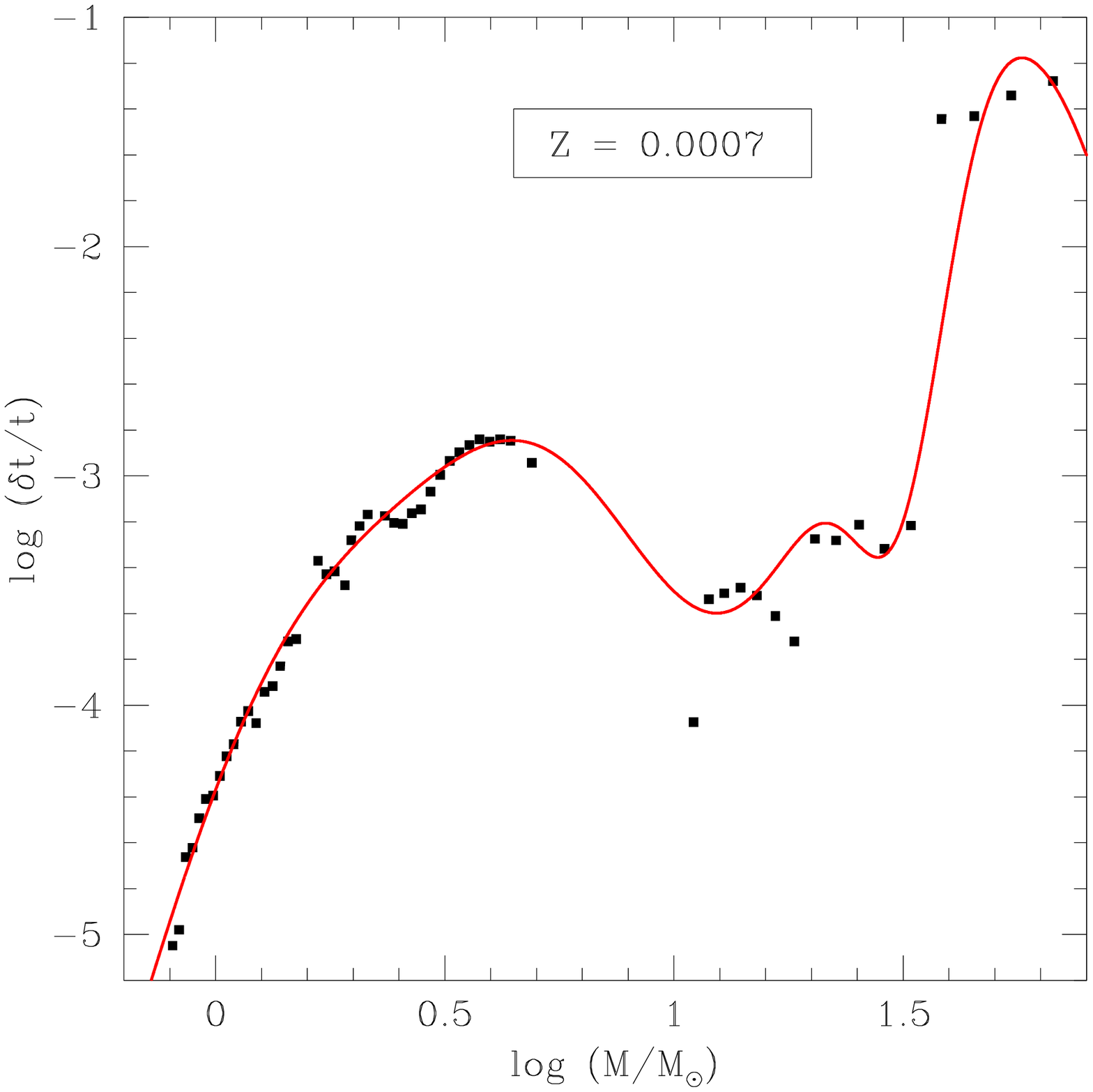,width=58mm}
}}
\caption[]{Same as figure~\ref{fig:fig12} but for $Z=0.0007$.}
\end{figure*}

\begin{figure*}
\centerline{\hbox{
\epsfig{figure=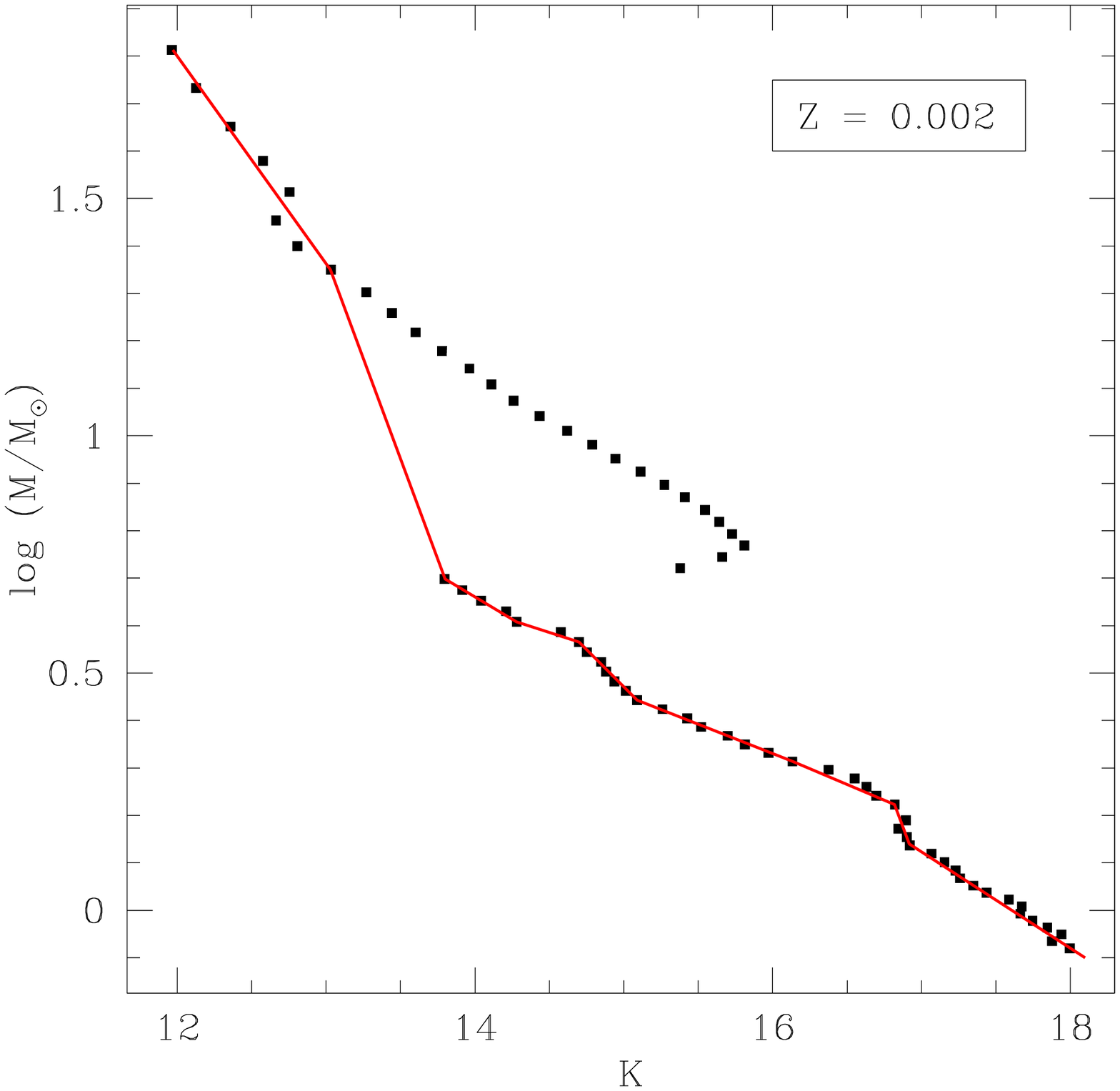,width=58mm}
\epsfig{figure=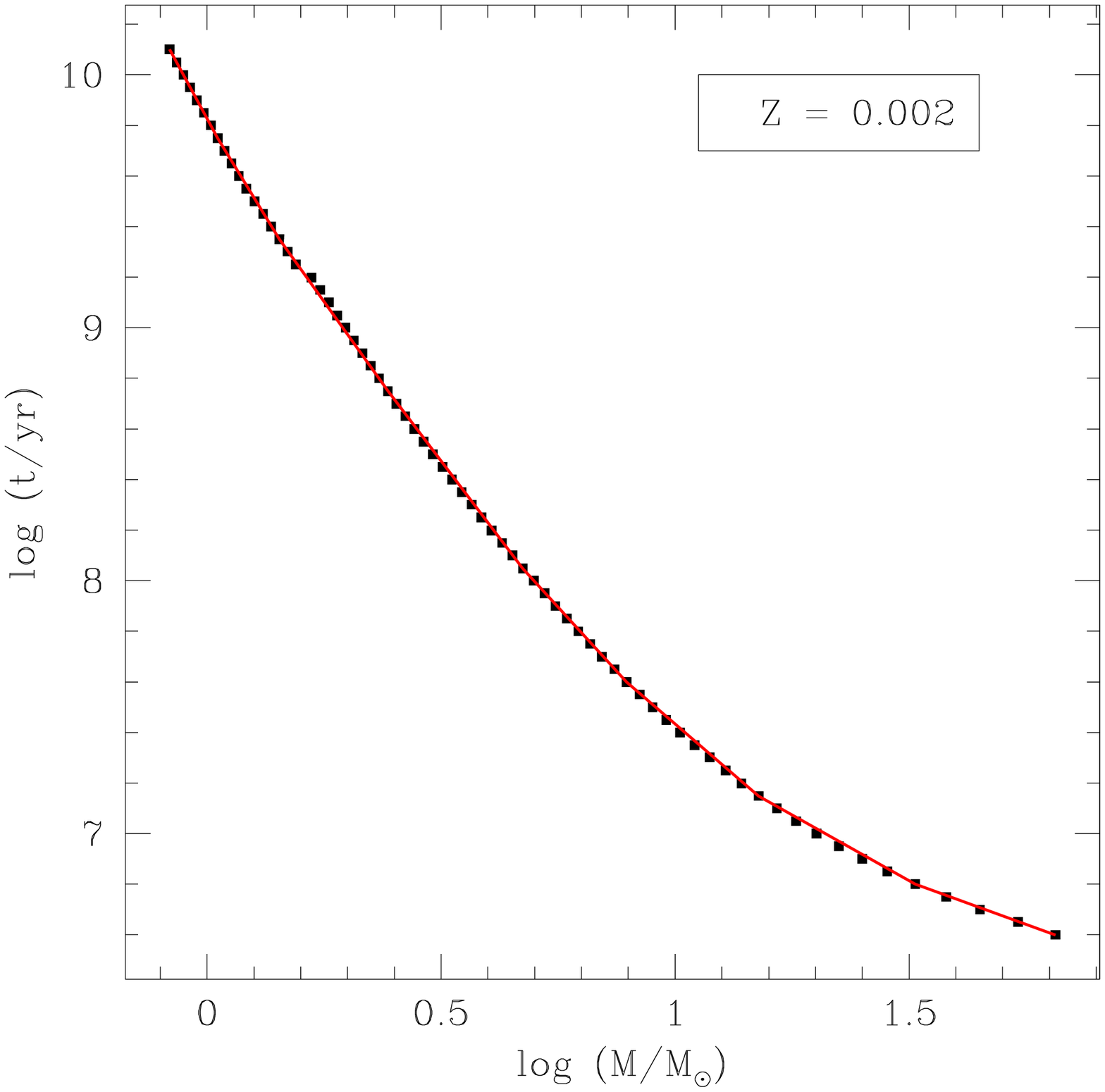,width=58mm}
\epsfig{figure=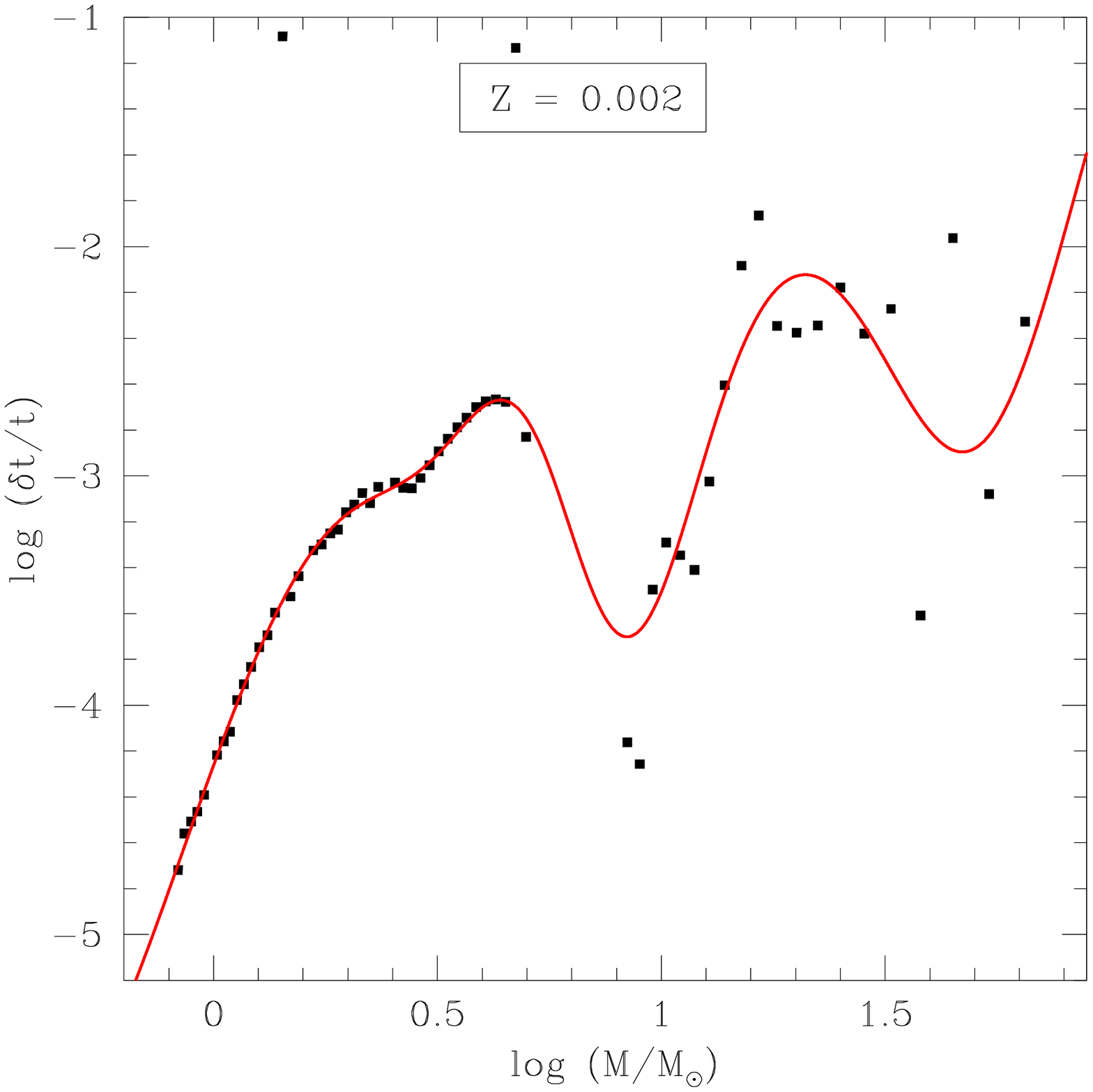,width=58mm}
}}
\caption[]{Same as figure~\ref{fig:fig12} but for $Z=0.002$.}
\end{figure*}

\begin{figure*}
\centerline{\hbox{
\epsfig{figure=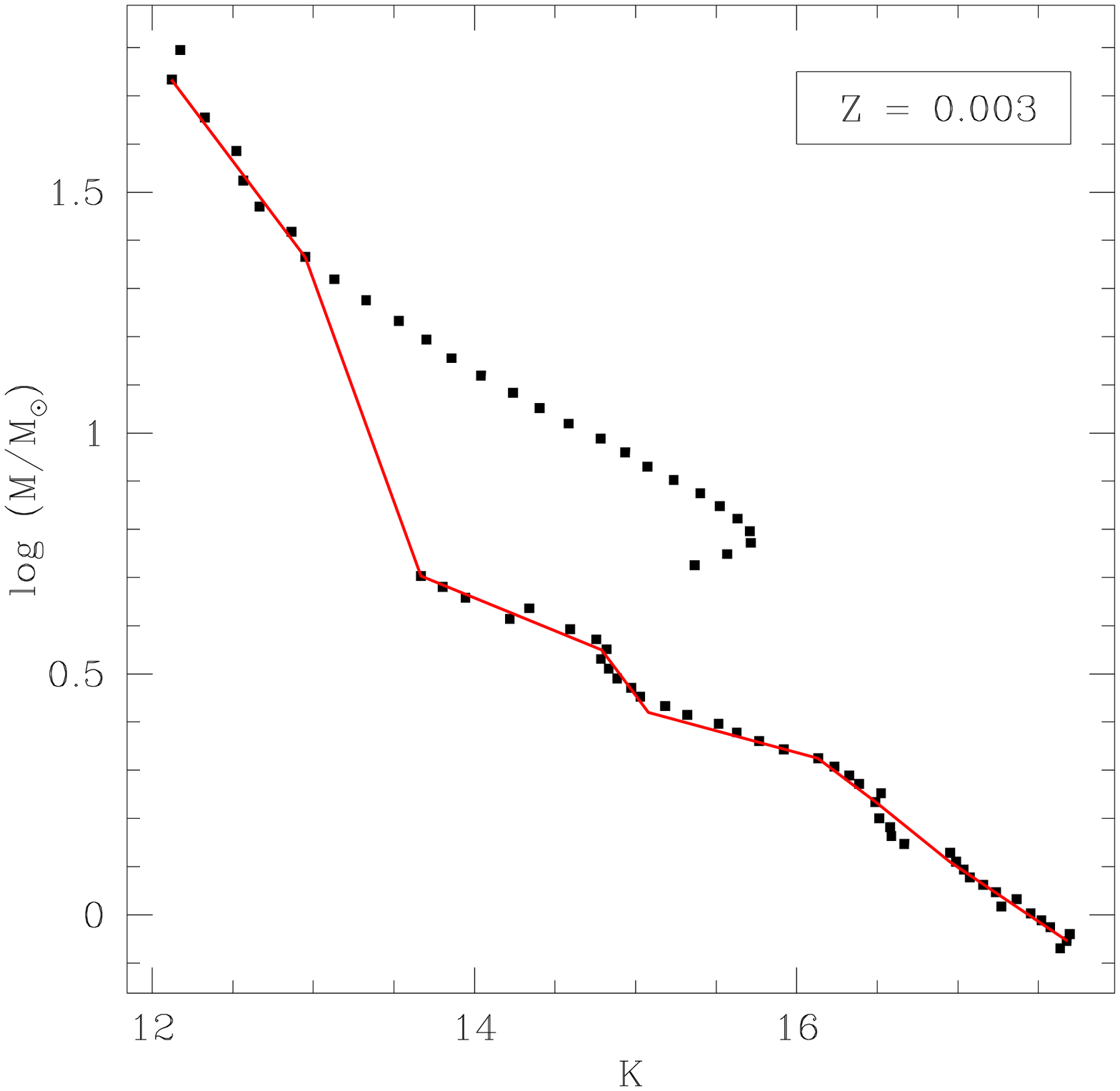,width=58mm}
\epsfig{figure=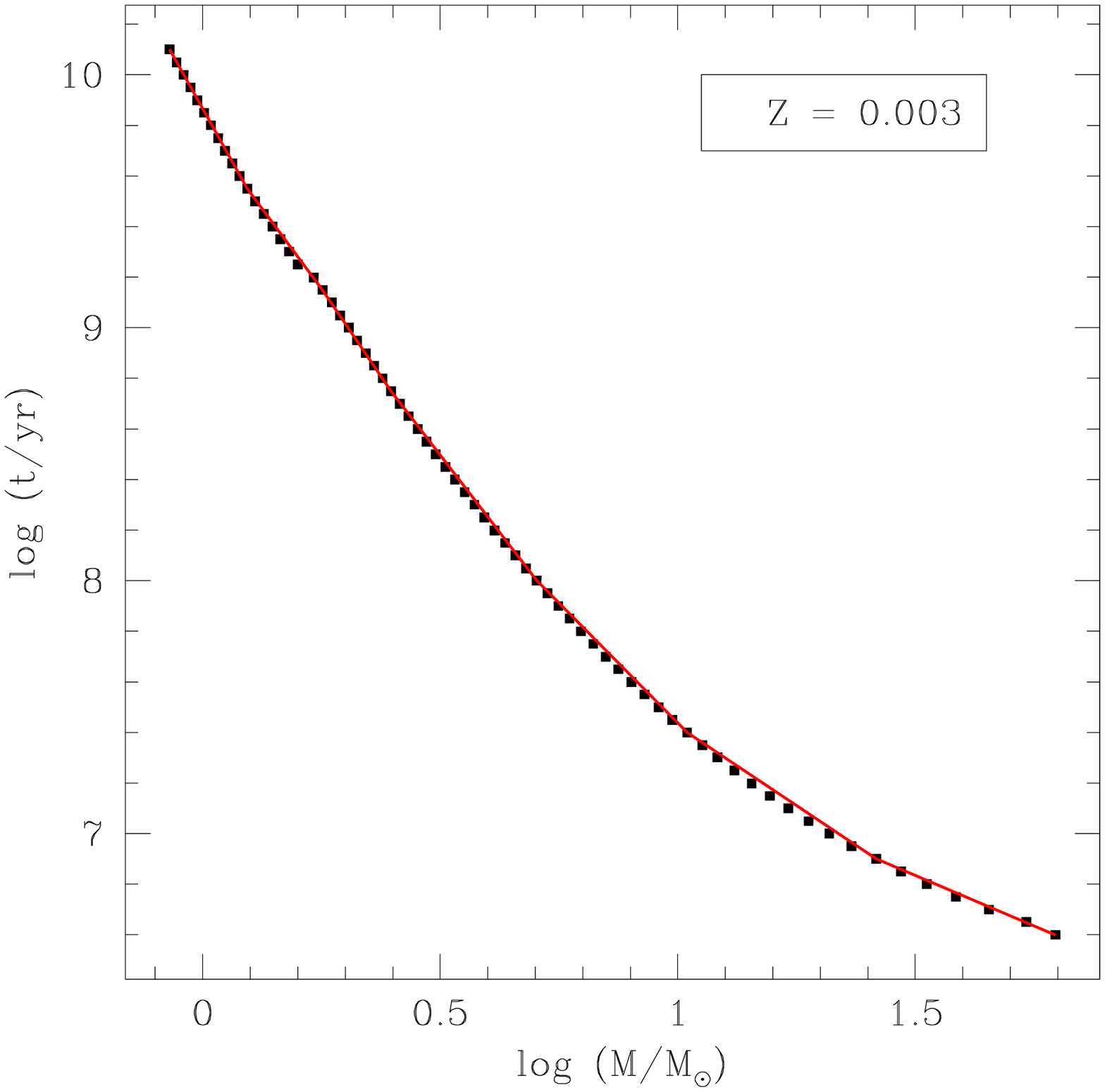,width=58mm}
\epsfig{figure=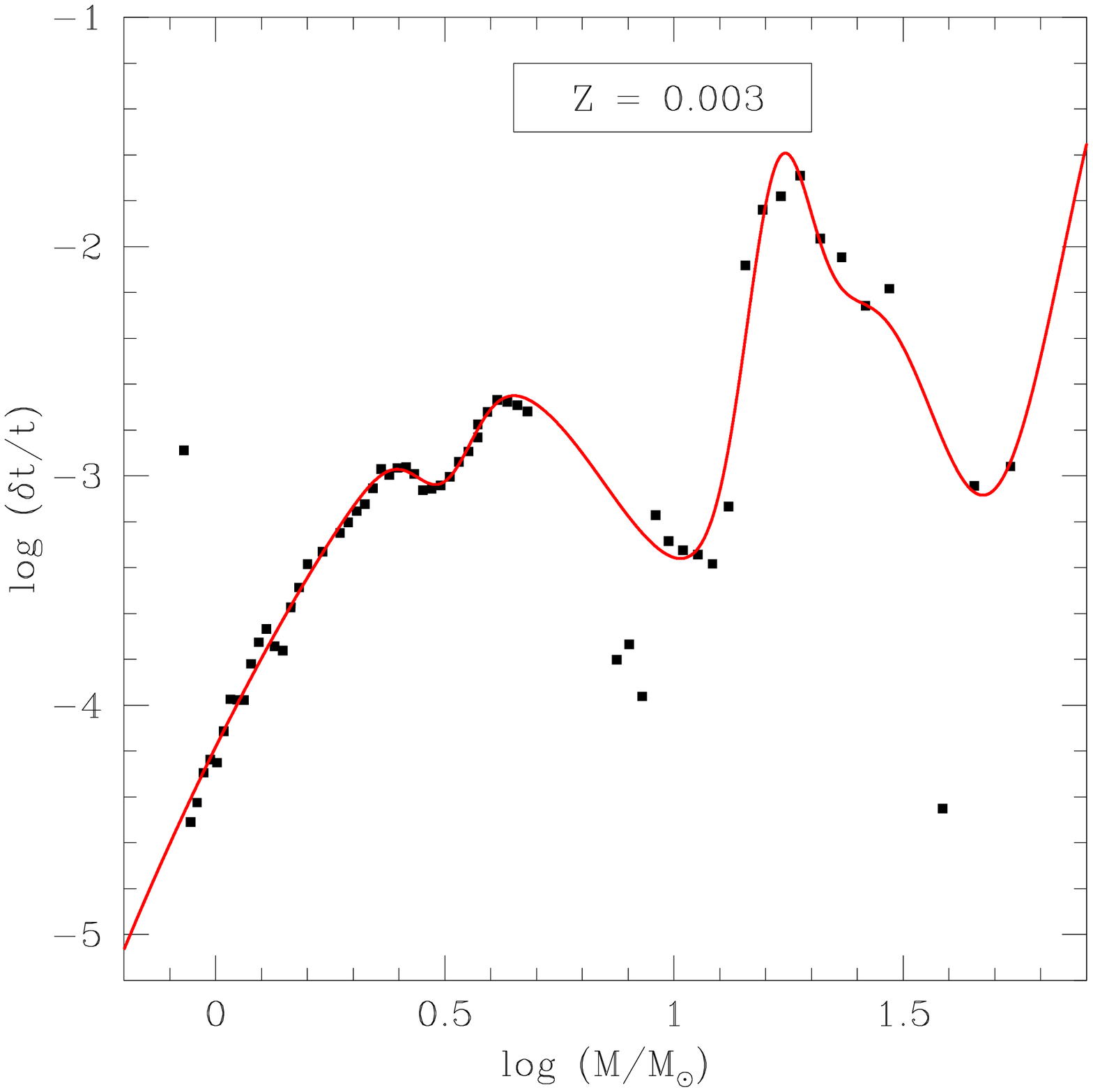,width=58mm}
}}
\caption[]{Same as figure~\ref{fig:fig12} but for $Z=0.003$.}
\end{figure*}

\clearpage

\begin{table}
\caption[]{Fits to the relation between birth mass and K-band magnitude, $\log
M/{\rm M}_\odot=aK+b$, for a distance modulus of $\mu=24.37$ mag.}
\begin{tabular}{ccr}
\hline\hline
       $a$        &       $b$       &   validity range   \\ 
\hline
\multicolumn{3}{c}{$Z=0.0004$} \\
\hline
$-0.263\pm0.008$  & $4.960\pm0.150$ & $K\leq13.52$       \\
$-8.248\pm0.246$  & $112.9\pm3.320$ & $13.52<K\leq13.61$ \\
$-0.102\pm0.003$  & $2.038\pm0.006$ & $13.61<K\leq14.92$ \\
$-0.374\pm0.011$  & $6.100\pm0.183$ & $14.92<K\leq15.35$ \\
$-0.010\pm0.003$  & $1.894\pm0.057$ & $15.35<K\leq16.19$ \\
$-0.1816\pm0.005$ & $3.219\pm0.096$ & $16.19<K\leq16.88$ \\
$-0.087\pm0.000$  & $1.627\pm0.048$ & $16.88<K\leq17.68$ \\
$-0.269\pm0.008$  & $4.839\pm0.144$ & $17.68<K\leq18.29$ \\
$-0.100\pm0.003$  & $1.749\pm0.051$ & $K>18.29$          \\
\hline
\multicolumn{3}{c}{$Z=0.0007$} \\
\hline
$-0.3875\pm0.011$ & $6.519\pm0.195$ & $K\leq13.20$       \\
$-1.882\pm0.057$  & $26.25\pm0.078$ & $13.20<K\leq13.59$ \\
$-0.105\pm0.003$  & $2.095\pm0.063$ & $13.59<K\leq14.70$ \\
$-0.393\pm0.012$  & $6.399\pm0.189$ & $14.70<K\leq15.12$ \\
$-0.116\pm0.004$  & $2.149\pm0.063$ & $15.12<K\leq16.38$ \\
$-0.214\pm0.006$  & $3.750\pm0.114$ & $16.38<K\leq16.93$ \\
$-0.040\pm0.001$  & $0.805\pm0.002$ & $16.93<K\leq17.37$ \\
$-0.231\pm0.007$  & $4.119\pm0.123$ & $K>17.37$          \\
\hline
\multicolumn{3}{c}{$Z=0.002$} \\
\hline
$-0.437\pm0.013$ & $7.041\pm0.210$ & $K\leq13.03$       \\
$-0.847\pm0.026$ & $12.38\pm0.360$ & $13.03<K\leq13.80$ \\
$-0.187\pm0.006$ & $3.280\pm0.099$ & $13.80<K\leq14.28$ \\
$-0.103\pm0.003$ & $2.074\pm0.063$ & $14.28<K\leq14.70$ \\
$-0.313\pm0.009$ & $5.171\pm0.156$ & $14.70<K\leq15.09$ \\
$-0.124\pm0.004$ & $2.315\pm0.069$ & $15.09<K\leq16.13$ \\
$-0.132\pm0.004$ & $2.444\pm0.072$ & $16.13<K\leq16.82$ \\
$-0.826\pm0.025$ & $14.15\pm0.420$ & $16.82<K\leq16.92$ \\
$-0.203\pm0.006$ & $3.581\pm0.108$ & $K>16.92$          \\
\hline
\multicolumn{3}{c}{$Z=0.003$} \\
\hline
$-0.445\pm0.013$  & $7.122\pm0.213$ & $K\leq12.95$       \\
$-0.920\pm0.003$  & $13.28\pm0.390$ & $12.95<K\leq13.67$ \\
$-0.136\pm0.004$  & $2.568\pm0.078$ & $13.67<K\leq14.79$ \\
$-0.448\pm0.014$  & $7.180\pm0.216$ & $14.79<K\leq15.08$ \\
$-0.0904\pm0.003$ & $1.783\pm0.054$ & $15.08<K\leq16.13$ \\
$-0.254\pm0.008$  & $4.429\pm0.132$ & $16.13<K\leq16.49$ \\
$-0.268\pm0.008$  & $4.646\pm0.138$ & $16.49<K\leq16.95$ \\
$-0.2265\pm0.007$ & $3.949\pm0.117$ & $K>16.95$          \\
\hline
\end{tabular}
\end{table}

\begin{table}
\caption[]{Fits to the relation between age and birth mass, $\log t=a\log
M/{\rm M}_\odot+b$.}
\begin{tabular}{ccr}
\hline \hline
      $a$       &      $b$       &        validity range           \\
\hline
\multicolumn{3}{c}{$Z=0.0004$} \\
\hline
$-3.333\pm0.099$ & $9.767\pm0.294$ & $\log{M/{\rm M}_\odot}\leq0.050$ \\
$-2.690\pm0.081$ & $9.734\pm0.291$ & $0.050<\log{M/{\rm M}_\odot}\leq0.236$ \\
$-2.680\pm0.081$ & $9.732\pm0.291$ & $0.236<\log{M/{\rm M}_\odot}\leq0.422$ \\
$-2.225\pm0.066$ & $9.540\pm0.285$ & $0.422<\log{M/{\rm M}_\odot}\leq0.759$ \\
$-1.732\pm0.051$ & $9.166\pm0.280$ & $0.759<\log{M/{\rm M}_\odot}\leq1.106$ \\
$-1.195\pm0.036$ & $8.571\pm0.260$ & $1.106<\log{M/{\rm M}_\odot}\leq1.399$ \\
$-0.778\pm0.023$ & $7.989\pm0.240$ & $\log{M/{\rm M}_\odot}>1.399$ \\
\hline
\multicolumn{3}{c}{$Z=0.0007$} \\
\hline
$-3.409\pm0.102$ & $9.780\pm0.294$ & $\log{M/{\rm M}_\odot}\leq0.010$ \\
$-3.191\pm0.096$ & $9.781\pm0.294$ & $0.010<\log{M/{\rm M}_\odot}\leq0.100$ \\
$-2.787\pm0.084$ & $9.745\pm0.291$ & $0.100<\log{M/{\rm M}_\odot}\leq0.196$ \\
$-2.586\pm0.078$ & $9.706\pm0.291$ & $0.196<\log{M/{\rm M}_\odot}\leq0.350$ \\
$-2.376\pm0.072$ & $9.632\pm0.288$ & $0.350<\log{M/{\rm M}_\odot}\leq0.660$ \\
$-1.940\pm0.057$ & $9.342\pm0.279$ & $0.660<\log{M/{\rm M}_\odot}\leq0.920$ \\
$-1.558\pm0.048$ & $8.989\pm0.270$ & $0.920<\log{M/{\rm M}_\odot}\leq1.181$ \\
$-1.076\pm0.033$ & $8.419\pm0.252$ & $1.181<\log{M/{\rm M}_\odot}\leq1.459$ \\
$-0.679\pm0.020$ & $7.841\pm0.234$ & $\log{M/{\rm M}_\odot}>1.459$ \\
\hline
\multicolumn{3}{c}{$Z=0.002$} \\
\hline
$-3.412\pm0.136$ & $9.827\pm0.294$ & $\log{M/{\rm M}_\odot}\leq0.020$ \\
$-3.041\pm0.090$ & $9.819\pm0.294$ & $0.020<\log{M/{\rm M}_\odot}\leq0.150$ \\
$-2.582\pm0.078$ & $9.748\pm0.291$ & $0.150<\log{M/{\rm M}_\odot}\leq0.380$ \\
$-2.431\pm0.096$ & $9.689\pm0.291$ & $0.380<\log{M/{\rm M}_\odot}\leq0.670$ \\
$-2.024\pm0,060$ & $9.415\pm0.282$ & $0.670<\log{M/{\rm M}_\odot}\leq0.890$ \\
$-1.598\pm0.048$ & $9.033\pm0.270$ & $0.890<\log{M/{\rm M}_\odot}\leq1.170$ \\
$-1.045\pm0.030$ & $8.381\pm0.252$ & $1.170<\log{M/{\rm M}_\odot}\leq1.510$ \\
$-0.667\pm0.020$ & $7.809\pm0.234$ & $\log{M/{\rm M}_\odot}>1.510$ \\
\hline
\multicolumn{3}{c}{$Z=0.003$} \\
\hline
$-3.360\pm0.102$ & $9.865\pm0.347$ & $\log{M/{\rm M}_\odot}\leq0.090$ \\
$-2.523\pm0.075$ & $9.786\pm0.030$ & $0.090<\log{M/{\rm M}_\odot}\leq0.250$ \\
$-2.769\pm0.084$ & $9.848\pm0.294$ & $0.250<\log{M/{\rm M}_\odot}\leq0.390$ \\
$-2.450\pm0.075$ & $9.722\pm0.291$ & $0.390<\log{M/{\rm M}_\odot}\leq0.700$ \\
$-1.892\pm0.060$ & $9.329\pm0.279$ & $0.700<\log{M/{\rm M}_\odot}\leq1.020$ \\
$-1.256\pm0.039$ & $8.681\pm0.261$ & $1.020<\log{M/{\rm M}_\odot}\leq1.410$ \\
$-0.796\pm0.024$ & $8.028\pm0.240$ & $\log{M/{\rm M}_\odot}>1.410$ \\
\hline
\end{tabular}
\end{table}

\begin{table}
\caption[]{Fits to the relation between the relative pulsation duration and
birth mass, $\log(\delta t/t)=\Sigma_{i=1}^5a_i\exp\left[-(\log M[{\rm
M}_\odot]-b_i)^2/c_i^2\right]$.}
\begin{tabular}{cccc}
\hline\hline
$i$ &       $a$       &       $b$       &  $c$  \\
\hline
\multicolumn{4}{c}{$Z=0.0004$} \\
\hline
 1  & \llap{$-$}113.2 & \llap{$-$}2.296 & 2.073 \\
 2  & \llap{$-$}1.084 &           0.969 & 0.239 \\
 3  &           55.97 & \llap{$-$}1.295 & 1.580 \\
 4  & \llap{$-$}1.963 &           1.456 & 0.248 \\
 5  &           0.000 &           2.385 & 0.022 \\
\hline
\multicolumn{4}{c}{$Z=0.0007$} \\
\hline
 1  & \llap{$-$}560.9 &           147.7 & 74.38 \\
 2  & \llap{$-$}2.084 &           0.468 & 0.404 \\
 3  &           15.58 &           1.161 & 1.313 \\
 4  & \llap{$-$}1.373 &           1.479 & 0.141 \\
 5  & \llap{$-$}7.463 &           1.132 & 0.531 \\
\hline
\multicolumn{4}{c}{$Z=0.002$} \\
\hline
 1  & \llap{$-$}2.525 &           0.741 & 0.924 \\
 2  &           9.158 &           0.735 & 0.238 \\
 3  & \llap{$-$}2.027 &           1.719 & 0.304 \\
 4  & \llap{$-$}9.475 &           0.754 & 0.258 \\
 5  & \llap{$-$}5.478 & \llap{$-$}0.489 & 0.606 \\
\hline
\multicolumn{4}{c}{$Z=0.003$} \\
\hline
 1  & \llap{$-$}721.0 & \llap{$-$}10.65 & 4.693 \\
 2  & \llap{$-$}0.340 &           0.499 & 0.090 \\
 3  & \llap{$-$}1.961 &           1.089 & 0.359 \\
 4  &           1.288 &           1.225 & 0.099 \\
 5  & \llap{$-$}2.240 &           1.697 & 0.223 \\
\hline
\end{tabular}
\end{table}

\label{lastpage}
\end{document}